\documentclass[pra,aps,10pt,showpacs,superscriptaddress,nofootinbib,twocolumn]{revtex4-1}

\usepackage[utf8]{inputenc}

\usepackage{latexsym,amsmath}
\usepackage{amsbsy}
\usepackage[pdftex]{graphicx}
\graphicspath{{./img/}}
\usepackage{amssymb}
\usepackage{epstopdf}
\usepackage[usenames]{color}
\usepackage{float}
\usepackage{hyperref}
\usepackage[normalem]{ulem}
\usepackage{physics}
\usepackage{chemformula}
\usepackage[inline]{enumitem}
\usepackage{braket}
\usepackage{natbib}
\usepackage{multirow}
\usepackage{subfigure}

\newcommand{\BigO}[2]{\mathcal{O}(#1^{#2})}
\newcommand{\BigOna}[1]{\BigO{n_{\mathrm{a}}}{#1}}
\newcommand{\FOIS}[2]{S_{#1}^{(#2)}}
\newcommand{\pert}[2]{\psi_{#1}^{(#2)}}
\newcommand{\pertNorm}[2]{N_{#1}^{(#2)}}
\newcommand{\pertEne}[2]{E_{#1}^{(#2)}}
\newcommand{\pertProj}[2]{P_{#1}^{(#2)}}

\begin{document}

\title{Quantum-centric strong and dynamical electron correlation: A  resource-efficient second-order $N$-electron valence perturbation theory formulation for near-term quantum devices
}

\author{Aaron Fitzpatrick}
\email{aaron.fitzpatrick@algorithmiq.fi}
\affiliation{Algorithmiq Ltd, Kanavakatu 3C, FI-00160 Helsinki, Finland}
\affiliation{Trinity Quantum Alliance, Unit 16, Trinity Technology and Enterprise Centre, Pearse Street, D02 YN67, Dublin 2, Ireland}

\author{N. Walter Talarico}
\email{walter.talarico@algorithmiq.fi}
\affiliation{Algorithmiq Ltd, Kanavakatu 3C, FI-00160 Helsinki, Finland}
\affiliation{HelTeq group, QTF Centre of Excellence, Department of Physics, University of Helsinki, P.O. Box 43, FI-00014 Helsinki, Finland.}

\author{Roberto Di Remigio Eikås}
\affiliation{Algorithmiq Ltd, Kanavakatu 3C, FI-00160 Helsinki, Finland}
\affiliation{Hylleraas Centre for Quantum Molecular Sciences, Department of Chemistry, UiT The
Arctic University of Norway, N-9037 Tromsø, Norway}

\author{Stefan Knecht}
\affiliation{Algorithmiq Ltd, Kanavakatu 3C, FI-00160 Helsinki, Finland}
%\affiliation{ETH Z{\"u}rich, Department of Chemistry and Applied Life Sciences,
%Vladimir-Prelog-Weg 1-5/10, CH-8093 Z{\"u}rich, Switzerland}

\begin{abstract}
We present a measurement-cost efficient implementation of Strongly-Contracted $N$-Electron Valence Perturbation Theory (SC-NEVPT2) for use on near-term quantum devices. At the heart of our algorithm we exploit the properties of adaptive Informationally Complete positive operator valued measures (IC-POVMs)
%first introduced in Refs.~  \citenum{garcia2021learning,nyka22a} 
to ``recycle" the measurement outcomes from a ground state energy estimation on a quantum device to reconstruct the matrix elements of the three- and four-body reduced density matrices for use in a subsequent CPU-driven NEVPT2 calculation. The proposed scheme is capable of producing results in good agreement with corresponding conventional NEVPT2 simulations, while significantly reducing the cost of quantum measurements and allowing for embarrassingly parallel estimations of higher-order RDMs in classical post-processing. Our scheme shows favourable scaling of the total number of shots with respect to system size. This paves the way for routine inclusion of dynamic electron correlation effects in hybrid quantum-classical computing pipelines.  
\end{abstract}

\maketitle
\section{Introduction}
In the past two decades, quantum chemistry experiments on quantum computers have successfully simulated the electronic structure and molecular properties of small chemical compounds such as, for example, \ch{H2}, \ch{LiH}, \ch{BeH2}, \ch{HeH+} as well as hydrogen chains containing up to 12 atoms  \cite{Lanyon2010,peru14a,Kandala2017,sche23c} and, more recently, we have even seen the dawn of simulating genuine reaction chemistry on near-term quantum devices \cite{OBrien2023,Liep24a}.
However, scaling quantum algorithms to be able to treat molecular systems of chemical relevance, such as for example for drug discovery and drug design,
remains a major challenge due to the limited availability, capability and stability, in particular noise resilience, of contemporary near-term quantum hardware.
Moreover, numerous computational experiments based on variational quantum algorithms have so far been restricted to active space sizes -- that is, tailored complete active  spaces (CAS) consisting of a number of active electrons and active molecular orbitals within which the quantum algorithm is supposed to provide a solution equivalent to full configuration interaction -- of the molecular species under consideration that often lead not only to an insufficient recovery of the electronic correlation energy beyond the starting mean-field picture, but also subsequently to an erroneous prediction of the molecules' chemical properties. 
A notable exception is thanks to recent efforts from Robledo-Moreno \emph{et al.} \cite{robl24a}\ who performed large-scale electronic-structure calculcations on contemporary near-term quantum hardware combining noisy quantum computer outcomes with massively parallel HPC resources. Although impressive on its own, the mixed quantum-HPC quantum experiments reported in Ref.~\citenum{robl24a}\ still required \emph{substantial} amounts of HPC resources to arrive at results that are well within reach of contemporary quantum chemical approaches using -- in comparison -- rather modest computational resources. 
Last, but not least, a surprisingly large multitude of quantum CAS calculations are often being carried out with the use of -- from a computational chemists' viewpoint -- unacceptably minimal atomic orbital basis sets \cite{Nagy17}.   

As pointed out above, the majority of present-day quantum algorithms set out to tackle the electronic structure problem within the so-called CAS formalism, ultimately leading to quantum simulations that are corresponding analogues of the CAS Configuration Interaction (CASCI) or CAS Self Consistent Field (CASSCF) approaches in ``traditional" \cite{note-traditional} quantum chemistry. 
While being among the most versatile methods in the toolbox of quantum chemists \cite{olse11,roos16a} and their ability to tackle many challenging problems in computational chemistry thanks to their multiconfigurational nature, CASCI and CASSCF by construction do not account for dynamical correlation particularly required for a \textit{quantitative} and sometimes even \textit{qualitative} correct description of the electronic structure in ground- and excited states of strongly correlated such as (multi-center) transition-metal complexes or extended aromatic compounds \cite{shee21}. The most popular quantum-chemical approaches for capturing such correlations based on a zeroth-order CAS(-like) reference wave function are the second order Perturbation Theory (CASPT2) \cite{Andersson1990,Roos1992} and N-electron Valence state second order Perturbation Theory (NEVPT2) \cite{ANGELI2001297,malrieu2001,jean-paul2002} approaches, extensive assessments of which can be found in \cite{Sarkar2022}.

Equipped with tools to identify and construct suitable CAS spaces in an automated way \cite{stei16a,stei19a,sayf17a}\ to drive \emph{resource-aware}\ and chemically-motivated zeroth-order quantum calculations, we present in this work a computational workflow for the inclusion of dynamical electron correlations from non-valence orbitals that combines 
our recently proposed self-consistent field adaptive variational quantum eigensolver (ADAPT-VQE-SCF) \cite{peru14a, grim19a, fitzpatrick2022selfconsistent}\ algorithm with the strongly contracted NEVPT2 (SC-NEVPT2) approach.
The original formulation of NEVPT2 in terms of spinfree reduced density matrices (RDMs) requires the calculation of three- and four-body RDMs within the active space \cite{jean-paul2002}, respectively, which entails a storage of $\BigOna{6}$ and $\BigOna{8}$ quantities, albeit lower-scaling variants have been proposed in the meantime \cite{Mahajan2019-va, Blunt2020-uj, Kollmar2021}. Hence, a quantum-centric NEVPT2 may quickly run the risk of facing an unsurmountable measurement overhead due to the large number of measurements required for each operator in the higher-order RDMs, in particular for near-term quantum devices. In contrast to earlier work \cite{Krompiec2022,Tammaro2023}, we propose to mitigate the measurement overhead with regard to the active space energy estimation by exploiting a recently introduced approach for energy evaluation relying on Adaptive Informationally complete generalised Measurements (AIM) \cite{garcia2021learning, glos22a}. Besides offering an efficient and scalable framework to measure the energy itself, Informationally Complete (IC) measurement data can be reused
to estimate all the matrix elements of the three and four-body RDMs for SC-NEVPT2, using only classical post-processing \cite{nyka22a}. 

The paper is organized as follows: We commence by giving a brief overview of the SC-NEVPT2 and IC-POVM theories in Sections \ref{sec:nevpt2}\ and \ref{sec:povm}, respectively, before presenting in Section \ref{sec:results}\ our quantum simulation results including PT2 energy corrections added to VQE-SCF energies by resorting to higher-order RDMs for the perturbation corrections that have been computed from both statevector (QASSCF-SV) and finite-statistics IC-POVM (QASSCF-POVM) data, respectively. In the latter case, we particularly highlight the fact that the IC-POVM-data driven higher-order RDMs have been obtained from adaptively optimised measurements for the (active-space) Hamiltonian of the target molecular system. Finally, we conclude with a discussion and perspective to make use of our proposed PT2 approach within quantum utility experiments for molecular systems on contemporary near-term quantum devices.

\begin{figure*}[ht!]
  \centering
  \includegraphics[width=6.5in]{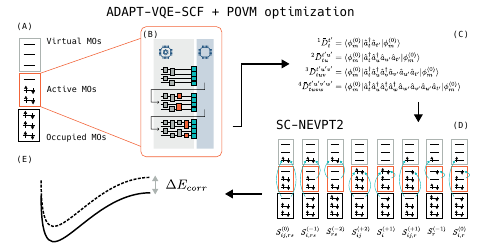}
     \caption{
     (\texttt{A}) Definition of the active space for reference multiconfigurational wavefunction, 
     (\texttt{B}) ADAPT-VQE-SCF optimization of the multiconfigurational reference, $\ket{\phi^{(0)}_m}$, with IC-POVM measurements,
     (\texttt{C}) Estimation of 1-, 2-, 3-, and 4-body RDMs with IC-POVM measurements,
     (\texttt{D}) Computation of SC-NEVPT2 energy contributions for different FOIS sectors, 
     (\texttt{E}) PES including static and dynamic correlation.}
  \label{fig:schematic}
\end{figure*}

\section{Second-order N-electron valence state perturbation theory}\label{sec:nevpt2}

In this section, we define the notation to be used throughout this work and briefly review the main features of NEVPT2, with particular focus on its strongly contracted (SC-NEVPT2) version and the computational bottleneck due to the fourth-order RDM appearing in the expressions of the second-order perturbative energy corrections. 

The starting point in multireference perturbation theory is the solution of the complete active space problem (CAS) in the form of a reference wave function:
\begin{equation}
\label{eq:ref_wf}
   \ket{\phi_m^{(0)}} = \sum_I C_{I,m} \ket{ D_I }
\end{equation}
where $m$ labels the state and $\ket{D_I}$ are configuration interaction (CI) determinants. Following standard conventions for the orbitals indices, we use $p, q, \ldots$, $i, j, \ldots$, $t, u, \ldots$, and $a, b, \ldots$ to denote general, doubly occupied (inactive), active, and virtual or secondary orbitals, respectively. 
In general and in particular when a balanced CAS has been chosen \cite{stei16b}, $\ket{ \phi_m^{(0)} }$ suffices to adequately capture static correlations present in the system and usually gives a satisfactory answer very close to the exact wave function. However, for quantitative agreement with the exact wave function, one must take into account dynamical electron correlation contributions by means of including various first-order interacting space (FOIS) excitations, that is inactive-secondary, inactive-active, active-secondary excitations as well as combinations thereof, see Fig.~\ref{fig:schematic}(D). The latter can be achieved by using second-order perturbation theory~\cite{Roos1982-th, Andersson1992-rc, Angeli2001, Angeli2001_2, Angeli2002-xw, Sokolov2016, Sokolov2017} that allows for perturbative inclusion of correlations involving inactive and secondary orbitals from an appropriate choice of a given reference Hamiltonian ($\hat{H_0}$). As a matter of fact, there is no unique way to define a reference Hamiltonian as the only requirement is that the reference wave function $\ket{\phi_m^{(0)}}$\ (see also Eq.~\eqref{eq:ref_wf}) is its ground state. Hence, different choices of the reference Hamiltonian lead to different perturbation theories~\cite{Pulay2011, Angeli2001}. In this work, we choose the Dyall Hamiltonian $\hat{H}^D$\ \cite{Dyall1995-ay} as the reference Hamiltonian $\hat{H}_0$\ which then forms the basis of the SC-NEVPT2 formalism.
In the strongly contracted scheme, the uncontracted FOIS is divided into subspaces $\FOIS{l}{k}$, where $k$ is the change in the number of active electrons relative to $\ket{\phi_m^{(0)}}$ ($-2 \le k \le 2$) and $l$ denotes the configuration of electrons in the core and virtual spaces. 
In the SC theory, only a single perturber state $\ket{\pert{l}{k}}$ is assigned to each class $\FOIS{l}{k}$\ where the active orbitals of the two particle excitations operators are contracted with parts of the perturbation operator leading to quasi-hole/particle energies in the denominator of the respective subspace second-order correction contributions to the energy \cite{jean-paul2002}. Specifically, 
\begin{equation}
\label{eq:perturber_wf}
   \ket{\pert{l}{k}} = \pertProj{l}{k}H \ket{\phi_m^{(0)}}
\end{equation}
where $\pertProj{l}{k} \equiv \ket{\pert{l}{k}}\bra{\pert{l}{k}}$ is the projector onto the $\FOIS{l}{k}$ subspace and $H$ is the Hamiltonian operator of the molecular system. In passing we note that the perturber functions $\ket{\pert{l}{k}}$\ in Eq.~\eqref{eq:perturber_wf} are orthogonal but are not normalized to unity, that is,
\begin{equation}
    \pertNorm{l}{k} =  \braket{\pert{l}{k} | \pert{l}{k}}\ ,
\end{equation}
are the squared norms of the perturbers $\ket{\pert{l}{k}}$. 
Given Eq.~\eqref{eq:perturber_wf}, the zeroth-order Hamiltonian for the SC-NEVPT2 is defined as 
\begin{equation}
   \hat{H}_0 = \sum_m E_m^{(0)} \ket{\phi_m^{(0)}}\bra{\phi_m^{(0)}}  +  \sum_{l,k} E_l^{(k)}  \ket{\pert{l}{k}} \bra{\pert{l}{k}}, 
\end{equation}
the second-order perturbative energy correction reads as \cite{Angeli2001}, 
\begin{equation}
\label{eq:2nd_energy}
    E_m^{(2)} = \sum_{l,k} \frac{\pertNorm{l}{j}}{E_m^{(0)} - \pertEne{l}{k} }.
\end{equation}

Here, $E_m^{(0)}$ is the zeroth-order energy for state $m$, and  $\pertEne{l}{k}$ are the perturber energies. In NEVPT2, these perturber energies are defined via the Dyall Hamiltonian $\hat{H}^D$, that is commonly chosen as the zeroth-order Hamiltonian:
\begin{equation}
   \pertEne{l}{k} = \frac{1}{\pertNorm{l}{k}} \braket{\pert{l}{k} |\hat{H}^{D}| \pert{l}{k}}
\end{equation}
with
\begin{equation}
    \hat{H}^D = \sum_i^{core} \epsilon_i \hat{a}_i^\dagger \hat{a}_i + \sum_a^{virtual} \epsilon_a \hat{a}_a^\dagger \hat{a}_a + \hat{H}_{active }, 
\end{equation}
where $\hat{H}_{active }$ is the core-averaged Hamiltonian in the active space such that $\hat{H}^{D}  | \phi_m^{(0)} \rangle= E_m^{(0)}  | \phi_m^{(0)} \rangle$. 

Hence, to compute the second-order energy by means of  Eq.~\eqref{eq:2nd_energy} requires the squared norms as well as the perturber energies. As elaborated in detail in the Appendix A of Ref.~\citenum{jean-paul2002}, calculating these quantities involves constructing higher-order active-space RDMs (with particle-number of up to four) and contraction with one and two electron integrals. In NEVPT2, the absence of interactions between subspaces allows the calculation of total correlation energies for each subspace separately. When using large active space references in NEVPT2 calculations, the internal and semi-external subspaces involving excitations in the active orbitals ($S_i^{(+1)}$, $S_r^{(-1)}$ in Fig.~\ref{fig:schematic}) become computational bottlenecks due to the need for three- and fourth-order RDM calculations, for example:

\begin{equation}
E_r^{(-1)} \propto\ 
\begin{array}{l} 
^3D_{tuv}^{t'u'v'} = \langle \phi_m^{(0)} | \hat{a}^\dagger_t \hat{a}^\dagger_u \hat{a}^\dagger_v \hat{a}_{v'} \hat{a}_{u'} \hat{a}_{t'} | \phi_m^{(0)} \rangle, \quad \\[3ex] ^4D_{tuvw}^{t'u'v'w'} = \langle \phi_m^{(0)} | \hat{a}^\dagger_t \hat{a}^\dagger_u \hat{a}^\dagger_v \hat{a}^\dagger_w \hat{a}_{w'} \hat{a}_{v'} \hat{a}_{u'} \hat{a}_{t'} | \phi_m^{(0)} \rangle
\end{array}
\end{equation}
whose measurement requirements scale as $\BigOna{6}$ and $\BigOna{8}$, respectively, in the number of active-space orbitals, $n_a$.

The cost of NEVPT2 calculations for large active spaces is dominated (i) by the solution of the CAS problem which scales exponentially with the size of the active space due to an exponential scaling of the CI basis and (ii) the requirement to compute in general higher-order RDMs which scale as high as $\BigOna{8}$.
By encoding the electronic occupation number vectors of length $N$ as a quantum register of $N$ qubits, the large CI basis, as well as the electronic Hamiltonian, can be expressed using the $2^N$ basis states of the $N$ qubits. 
Quantum algorithms like quantum phase estimation (QPE)~\cite{kita95a, guth20a} or variational quantum eigensolver (VQE) \cite{peru14a, till22a, fedo22a} have the potential to solve the electronic structure problem requiring a number of steps only polynomial in $N$, suggesting an exponential speedup with respect to the classical approach. 
Thus, by substituting the CAS-CI solver with an efficient quantum algorithm (such as the VQE), it is possible to further enhance the realm of multireference perturbation theory techniques to larger active spaces and more complex chemical systems. Such implementations have recently been reported, in which NEVPT2 correction is carried out at different level of approximation ~\cite{Krompiec2022, Tammaro2023, Gunther2023}. 

For example, \citet{Krompiec2022} suggested a new approximation to the 4-RDM, inspired by and simultaneously addressing the shortcomings of the cumulant approximation to higher-order RDMs \cite{kutz10a}. While the latter reduces the number of matrix elements to evaluate to arrive at a complete higher-order RDM, it still results in an overall increase in the number of measurement circuits in comparison to energy-only measurements. \citet{Tammaro2023} make use of a VQE ansatz together with the quantum subspace expansion algorithm (QSE)~\cite{mcclean2017} to evaluate the eigenvectors and eigenvalues of the Dyall Hamiltonian that are then used for the second-order energy correction. Notably, they conclude that the  accuracy of the obtained results depends on how many excited states are included and this again could generally lead to an increase in the number of measurements. 

In this work, we take a similar approach as in Refs.~\citenum{Krompiec2022} and \citenum{Tammaro2023}, respectively, but we replace the CAS-CI component with an ADAPT-VQE-SCF implementation ~\cite{peru14a, grim19a, fitzpatrick2022selfconsistent} together with informationally complete measurements~\cite{garcia2021learning, glos22a, nyka22a} for the estimation of RDM matrix elements. This scheme allows us to benefit from the quantum advantage of the state preparation step while offloading the measurements overhead of the RDM matrix elements in a classical post-processing as explained in the next section. Most importantly, neither approximations to higher-order RDMs are required nor does our proposed algorithm incur any measurement overhead.

\section{Adaptive informationally complete
measurements}\label{sec:povm}
In standard quantum computing algorithms, measurement by projection into the computational basis is very expensive in terms of the total number of measurement rounds required to reach a good accuracy for the estimation of a given observable. Even if one's interest lies in determining the average of a single operator, such as the system's energy $\braket{H}$, the scaling of the necessary measurements quickly increases with the size of the system, particularly for operators that, when mapped into qubit basis, involve large linear combinations of Pauli strings. 
The excessive number of measurements required for any VQE-based algorithm has been identified as a potential roadblock for any real-world applications~\cite{Gonthier2022}. To address this challenge, several approaches aimed at minimizing the required measurements for the estimation of physically relevant observables have been put forward~\cite{Kandala2017,Izmaylov2019,Zhao2020, Huggins2021, Crawford2021,Zhao2020,Verteletskyi2020}. 
\citet{garcia2021learning} proposed using adaptive informationally complete (IC) generalized measurements to enhance the measurement scheme on-the-fly during the computation~\cite{garcia2021learning}. 
The generalized measurement scheme, which for completeness is briefly described later, has been demonstrated to
\begin{enumerate*}[label={\emph{\alph*})}]
    \item significantly decrease the cost of energy estimation, and
    \item utilize the measurement data for estimating expectation values of other operators, regardless of whether they commute with the Hamiltonian.
\end{enumerate*}  
Significantly, various implementations of Informationally Complete Positive Operator-Valued Measures (IC-POVMs) have been suggested for gate-based quantum computers. These include dilation implementations that make use of auxiliary qubits~\cite{garcia2021learning} or take advantage of the physical qubits higher energy levels~\cite{Fischer2022}, or dilation-free schemes that use randomized projective measurements~\cite{glos22a}. In this work, we consider the first of these implementations, in particular two different flavors that arise from two distinct parametrizations, and compare their performance when used to measure higher-order RDM for subsequent use in the NEVPT2 approach. We note, due to constraints on quantum hardware resources, that the dilation-free schemes should be utilised on a hardware implementation of the following calculations. This allows for the same functionality as presented here without the need of doubling the qubit resource requirements.

\begin{figure*}[ht!]
  \centering
  \includegraphics[width=.5\textwidth]{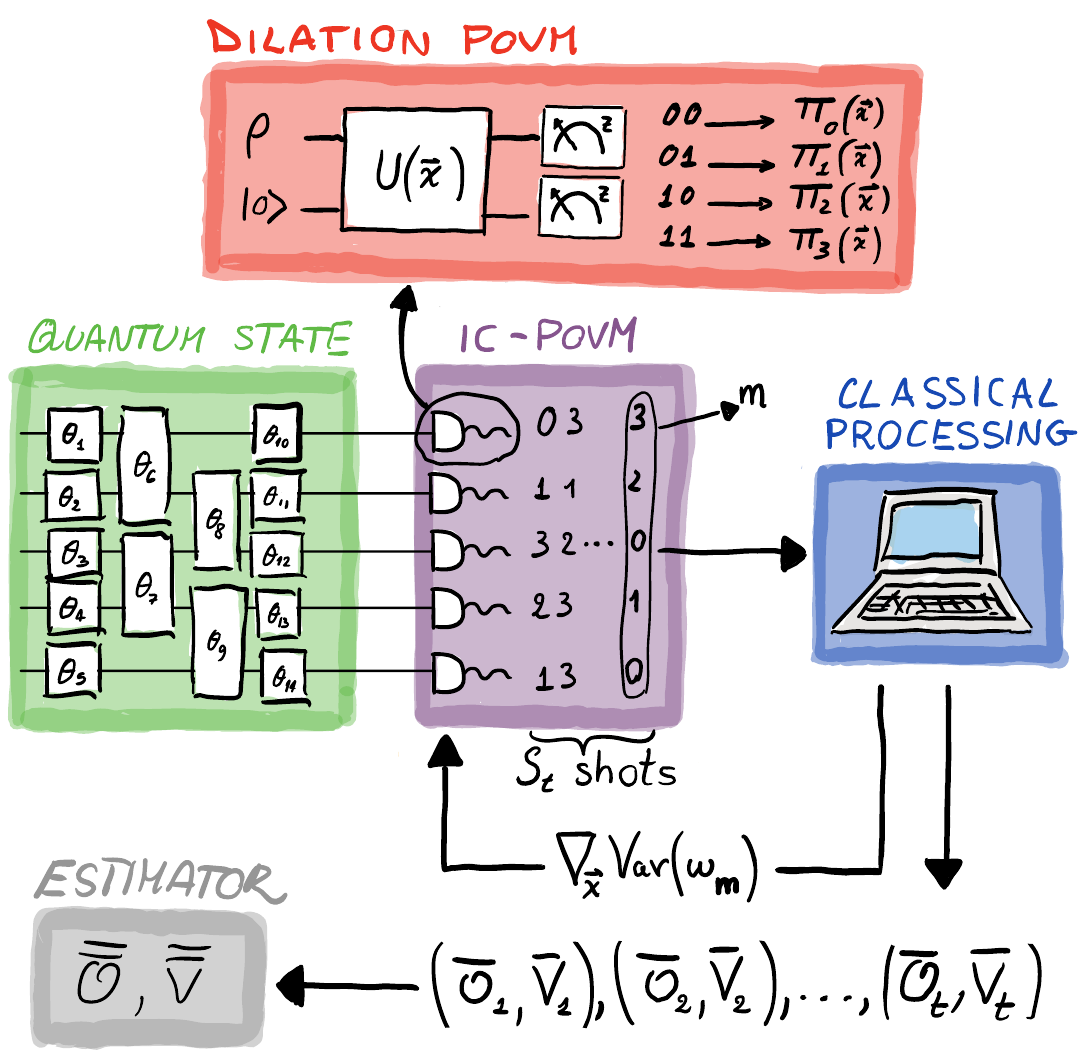}
  \caption{Adaptive measurement strategy based on IC-POVMs which allows to optimize the measurement (minimize the variance) for any given operator while being able to compute expectation values for any other operator in postprocessing. Reproduced from: García-Pérez, G.; Rossi, M. A. C.; Sokolov, B.; Tacchino, F.; Barkoutsos, P. K.; Mazzola, G.; Tavernelli, I.; Maniscalco, S. \emph{Learning to Measure: Adaptive Informationally Complete Generalized Measurements for Quantum Algorithms}. \textbf{PRX Quantum} 2021, 2 (4), 040342. \url{https://doi.org/10.1103/PRXQuantum.2.040342}. For further information, see text.}
  \label{fig:adaptive-ic-povm}
\end{figure*}

\subsection{Dilation POVM}
For a single qubit $i$, the POVMs are generalised quantum measurements described by four POVM effects $\{\Pi_{m_i}\}$, $m_i = 0, \dots, 3$ that are positive semi-definite operators $\Pi_{m_i}$ adding up to identity $\sum_{m_i} \Pi_{m_i} = \mathbb{I}$. This means that if a measurement is performed on a qubit in state $\rho^{(i)}$, the probability of obtaining outcome $m_i$ is given by $p_{m_i} = \mathrm{Tr} [  \rho^{(i)} \Pi_{m_i} ]$. 
If the effects are linearly independent, thus spanning the space of linear operators in the Hilbert space of the qubit, the corresponding POVM is informationally complete. Any one-qubit operator can be decomposed in the basis of the POVM effects. 
If we have an $N$-qubit system and every qubit is measured with a local POVM, the probability to obtain an outcome $\mathbf{m} = (m_1, \dots , m_N )$ on a given state $\rho$ with $m_i \in \{0, 1, 2, 3\}$, is $\mathrm{Tr}[\rho \Pi_{\mathbf{m}}]$, where $\Pi_\mathbf{m} = \bigotimes_{i=1}^N \Pi_{m_i}^{(i)}$. Notably, if each of the single-qubit POVMs is IC, the set of $4^N$ effects $\{ \Pi_\mathbf{m} \}$ is also IC in the space of linear operators of the $N$-qubit Hilbert space, $\mathcal{H}^{\otimes N}$. 
The effects $\{ \Pi_\mathbf{m} \}$ form a basis: any observable $O$ can be decomposed as $O = \sum_\mathbf{m} \omega_\mathbf{m} \Pi_\mathbf{m}$. Its expectation value can be estimated with measurement data obtained from these POVMs as follows:

\begin{equation}
\label{eq:exp_val}
\langle O \rangle = \mathrm{Tr} [\rho O] = \sum_\mathbf{m} \omega_\mathbf{m} \mathrm{Tr} [\rho \Pi_\mathbf{m}] = \sum_\mathbf{m} \omega_\mathbf{m} p_\mathbf{m}
\end{equation}
where $p_\mathbf{m}$ is the probability of obtaining outcome $\mathbf{m}$. Thus, the estimation can be carried out stochastically via a Monte Carlo approach: in particular, one can use quantum computation to sample from the probability distribution $\{p_\mathbf{m}\}$ and average the $\omega_\mathbf{m}$ corresponding to each result. The measurement is then repeated $S$ times to obtain different measurement outcomes $\mathbf{m}_1, \dots,  \mathbf{m}_S$ and the expectation value of an operator can be expressed accordingly as the sample mean over the different realizations:
\begin{equation}
\label{eq:_est_exp_val}
\bar{O} = \frac{1}{S} \sum_{s=1}^N \omega_{\mathbf{m}_s}
\end{equation}

Naturally, the accuracy for the estimated measurement is affected by stochastic fluctuations and the error in the estimation can be quantified by the standard error of the sample variance of the mean:
\begin{equation}
   \sigma_{\bar{O}} = \sqrt{\frac{\mathrm{Var}( \bar{O}) }{S}}  = \sqrt{ \frac{1}{S} \left( \langle \omega_\mathbf{m}^2 \rangle_{\{ p_\mathbf{m} \}} - \langle \omega_\mathbf{m} \rangle_{\{ p_\mathbf{m} \}}^2 \right)  },
\end{equation}
this depends on the probability distribution $\{p_\mathbf{m}\}$ as well as on the operator decomposition $\omega_\mathbf{m}$ in the basis of the POVM effects. 
In order to lower the standard error of a specific measurement, it is possible to parameterize and optimize the POVM effects such that the probability distribution is tailored to the problem at hand. This helps identifying a measurement to decrease the variance of the estimation, leading to an increasingly precise evaluation. We summarize in a schematic representation in  Figure~\ref{fig:adaptive-ic-povm} our adaptive measurement strategy.
POVM optimization can be achieved by \emph{reprocessing} the same measurement data used for the estimation of a specific observable. 
More specifically, measurements are carried out in small batches with IC data used to adjust the parameterized POVM effects and estimate the desired observables on-the-fly during the self-consistent optimization as depicted in Fig.~\ref{fig:n2-restricted}.
The process is repeated up until a desired standard error is reached and the number of shots in each round is updated accordingly so that the measurements are performed more efficiently~\cite{garcia2021learning}. 
Using this scheme, one can estimate different observables with the same measurement data while simultaneously mitigating the measurement overhead in quantum algorithms that require the evaluations of additional observables~\cite{nyka22a, fitzpatrick2022selfconsistent}. In the following, the parameterised IC-POVMs have been optimized to reduce the standard error in the Hamiltonian operator, though one could consider optimizing for, \emph{e.g.}, the number operator, which would lead to different and potentially better estimations of the RDM matrix elements.

\section{Results}\label{sec:results}
To demonstrate the validity and potential of our proposed IC quantum-centric NEVPT2 scheme, we study in the following perturbative energy corrections for several molecular systems. 
We assess the accuracy of our approach by considering noiseless simulations which we  compare with the corresponding classical CASSCF+SC-NEVPT2 calculations. Moreover, we further consider the effect of shot-noise arising from the statistical nature of the POVM measurement scheme and intrinsic to any quantum-centric approach that is fundamentally limited by statistical noise.

Mean-field orbitals and molecular integrals are generated in a pre-processing stage using the \textsc{PySCF}~\cite{sunq17b,sunq20a} quantum chemistry program suite on classical computers. These data is used as input to perform computations with quantum simulators, using our \textsc{Aurora}~software framework~\cite{aurora}. \textsc{Aurora} generates and optimizes active-space wavefunction ansatze \emph{via} ADAPT-VQE-SCF \cite{fitzpatrick2022selfconsistent} and estimates the RDMs \emph{via} POVM sampling for the subsequent SC-NEVPT2 calculations, again implemented as a classical post-processing step in the \textsc{PySCF}~program.\cite{Sokolov2016,Sokolov2017}

\subsection{Diatomic molecules}\label{sec:results:diatomics}
As a simple first example, we consider the ground-state potential energy curve (PEC) of \ch{LiH}.
The active space comprises four electrons in four orbitals, that is a CAS$(4e,4o)$ and employing a correlation consistent basis set, cc-pVDZ for both Li and H \cite{dunn89}, such that we are considering a significant number of virtual orbitals needed for a good estimation of the perturbative energy correction, while also easily being able to classically compute the FCI ground state. For the sampling scheme we test two distinct parametrizations of the POVM effects with different numbers of free parameters: 
\begin{enumerate*}[label=\emph{(\arabic*)}] 
\item an 8-parameter class (8P-POVM), where we optimize all the 8 degrees of freedom that are the minimum requirement to fully describe the space of single-qubit IC dilation POVMs~\cite{garcia2021learning}, and 
\item a 4-parameter class (4P-POVM) with fewer degrees of freedom, where only 4 parameters, representing geometric transformations of symmetric POVM on the Bloch sphere, are left free to vary. Three of these parameters define the rotation of the symmetric tetrahedron POVM in the Bloch sphere, whereas the remaining one defines the stretch of the POVM. The stretch describes how much  the tetrahedron is transformed into a projector, for which one of the effects is found on the surface and the remaining effects in the center of the Bloch sphere.
\end{enumerate*}
Since both of these two parametrizations belong to the same family of dilation POVMs, we do not expect a considerable difference in the accuracy of the estimations that they can perform, rather a distinction in the efficiency of the optimization. Indeed, we anticipate the 4P-POVM to be more resource-effective, in terms of measurements rounds,  compared to the 8P-POVM as its optimization is carried out in a restricted parameter space.

\begin{figure}[H]
  \subfigure[
  Requested threshold error $\sigma = 1.6\times 10^{-3}\,\mathrm{Ha}$ for 8P-POVM measurement estimation of the ground state energy.\label{fig:lih-dilation-left}]{%
    \includegraphics[width=0.45\textwidth]{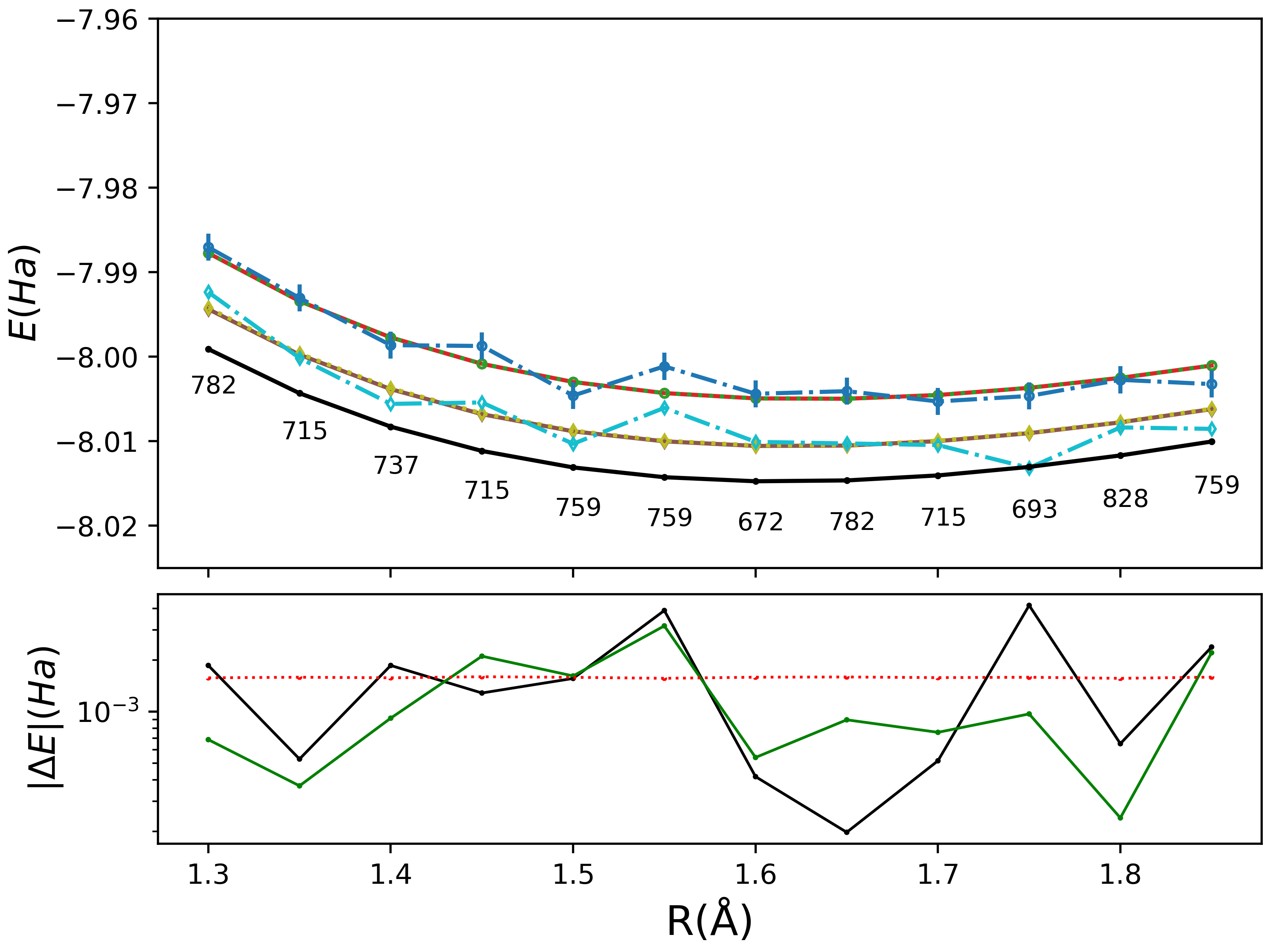}
  }
  \hfill
  \subfigure[
  Requested threshold error $\sigma = 0.8\times 10^{-3}\,\mathrm{Ha}$ for 8P-POVM measurement estimation of the ground state energy.\label{fig:lih-dilation-right}]{%
    \includegraphics[width=0.45\textwidth]{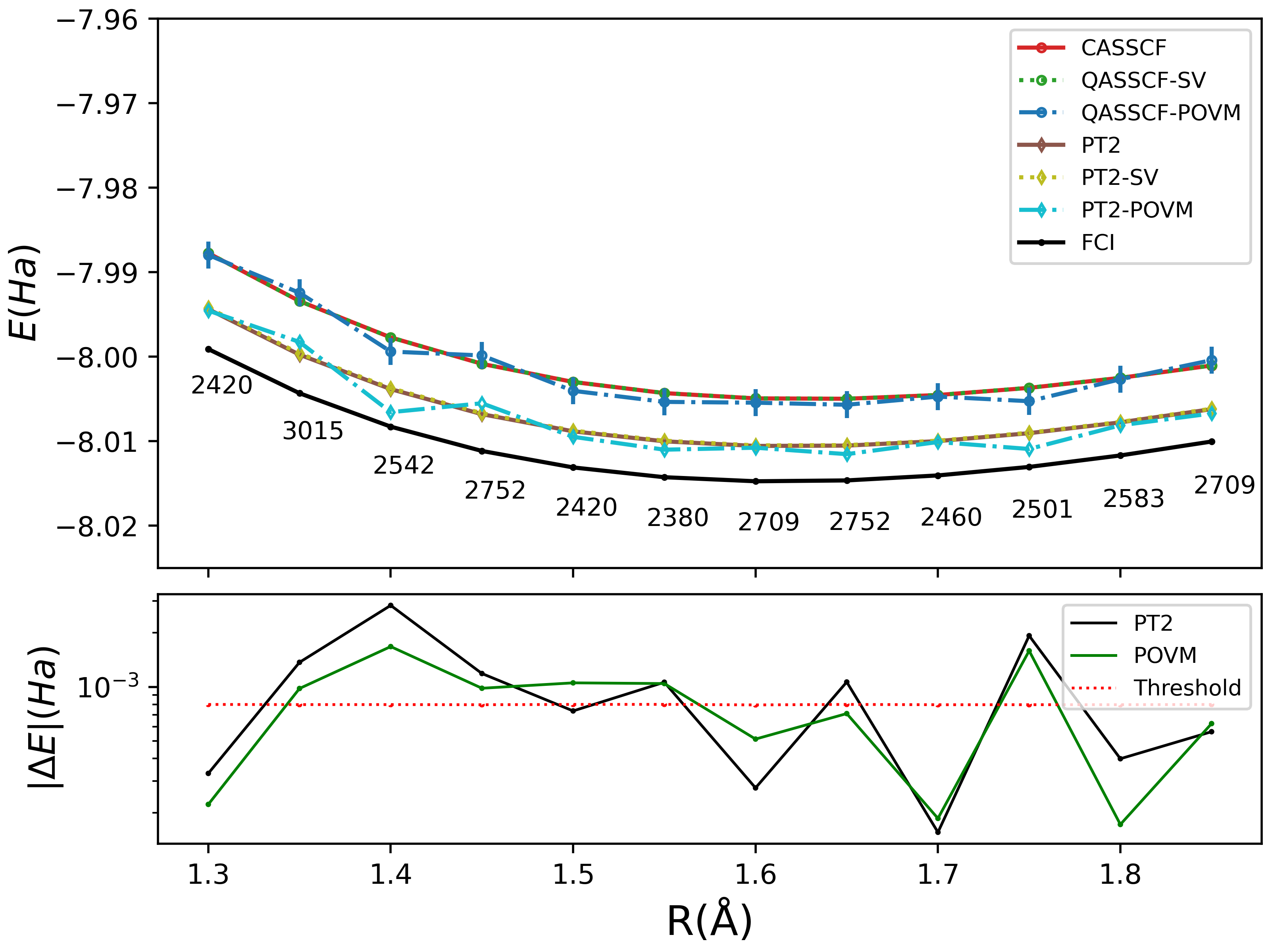}
  }
  \caption{Potential energy curves for \ch{LiH} \emph{via} optimized 8-parameter POVM for different requested threshold errors: left panel, $\sigma = 1.6\times 10^{-3}\,\mathrm{Ha}$, right panel, $\sigma = 0.8\times 10^{-3}\,\mathrm{Ha}$.  
  \textbf{Top} Converged noiseless simulation QASSCF-SV (green, dotted) and shots-noise estimation QASSCF-POVM (blue, dash-dotted) together with the estimated perturbative energy correction PT2-POVM (cyan, dash-dotted) compared with the classical reference CASSCF (red, solid), PT2 (brown, solid) and FCI (black, solid) calculations. The numbers below the FCI curves show the total amount of shots required (in 1000's) for the 8P-POVM measurement to estimate the ground state energy at given threshold error. \textbf{Bottom} Absolute energy difference between the classical calculation and the POVM estimation for the converged energy (green) and the perturbative energy correction (black) together with the relative threshold error (red).}
  \label{fig:lih-dilation}
\end{figure}

\begin{figure}[H]
  \subfigure[
  Requested threshold error $\sigma = 1.6\times 10^{-3}\,\mathrm{Ha}$ for 4P-POVM measurement estimation of the ground state energy.\label{fig:lih-restricted-left}]{%
    \includegraphics[width=0.45\textwidth]{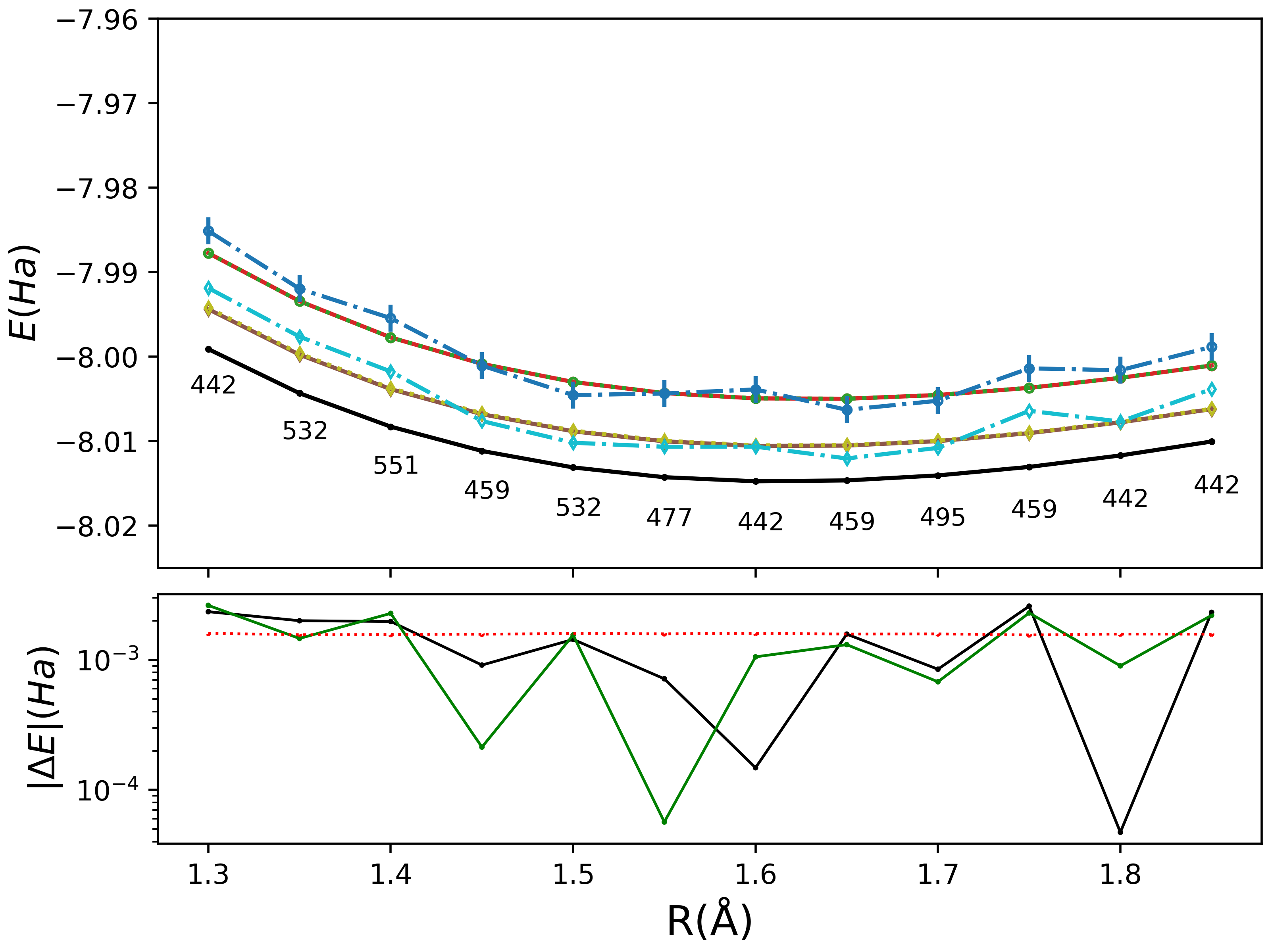}
  }
  \hfill
  \subfigure[
  Requested threshold error $\sigma = 0.8\times 10^{-3}\,\mathrm{Ha}$ for 4P-POVM measurement estimation of the ground state energy.\label{fig:lih-restricted-right}]{%
    \includegraphics[width=0.45\textwidth]{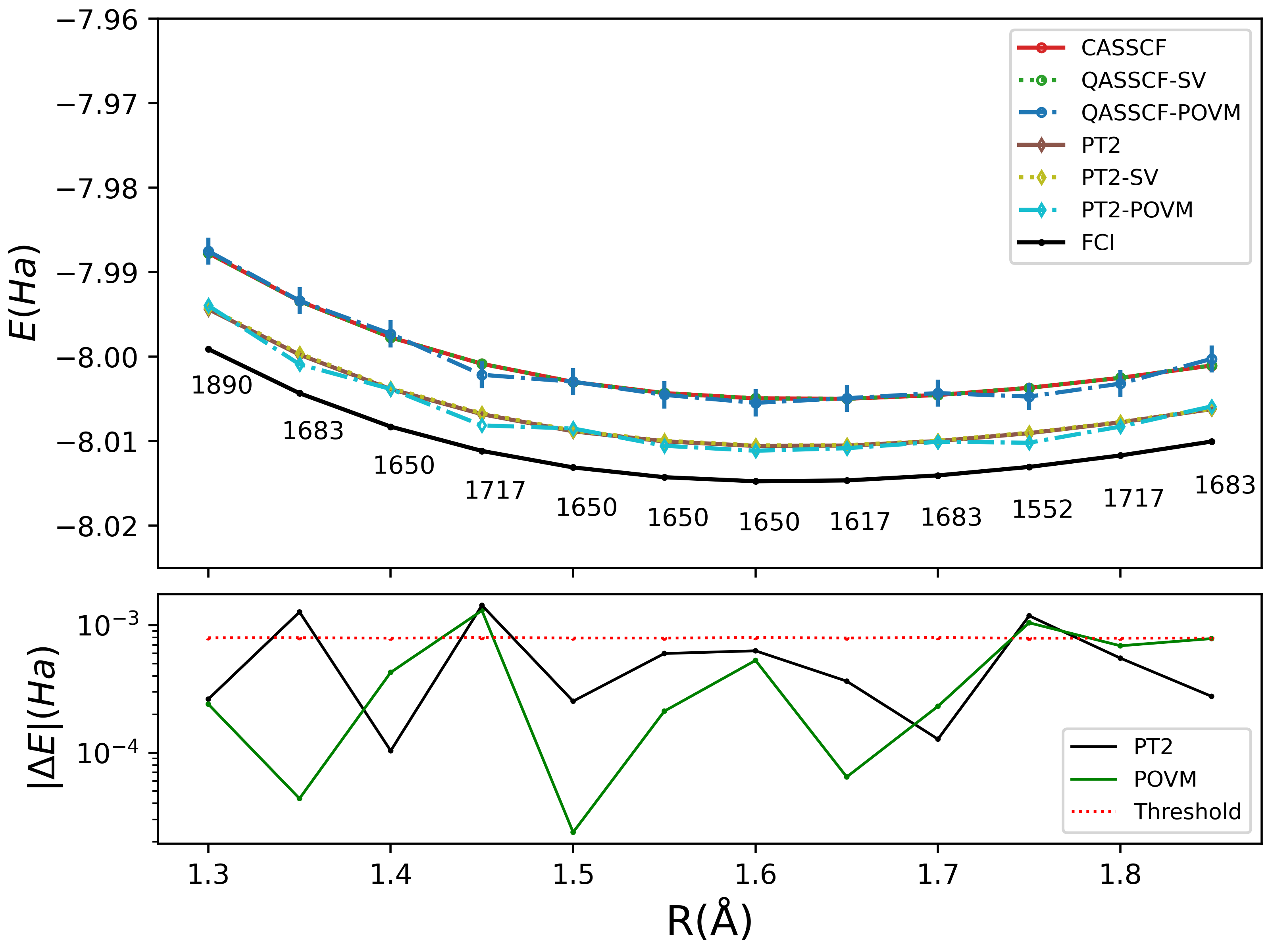}
  }
  \caption{Potential energy curves for \ch{LiH} \emph{via} optimized 4-parameter POVM for different requested threshold errors: left panel, $\sigma = 1.6\times 10^{-3}\,\mathrm{Ha}$, right panel, $\sigma = 0.8\times 10^{-3}\,\mathrm{Ha}$.
  \textbf{Top} Converged noiseless simulation QASSCF-SV (green, dotted) and shots-noise estimation QASSCF-POVM (blue, dash-dotted) together with the estimated perturbative energy correction PT2-POVM (cyan, dash-dotted) compared with the classical reference CASSCF (red, solid), PT2 (brown, solid) and FCI (black, solid) calculations. The numbers below the FCI curves show the total amount of shots required (in 1000's) for the 4P-POVM measurement to estimate the ground state energy at given threshold error. \textbf{Bottom} Absolute energy difference between the classical calculation and the POVM estimation for the converged energy (green) and the perturbative energy correction (black) together with the relative threshold error (red).}
  \label{fig:lih-restricted}
\end{figure}

Figures~\ref{fig:lih-dilation} and~\ref{fig:lih-restricted} show the second-order perturbative energy correction estimated via optimized 8P-POVM and 4P-POVM, respectively. The top panels show the PECs for $\ch{LiH}$ from the converged noiseless simulation within the QASSCF-SV scheme (green dotted lines) and the shots-noise estimation via POVM sampling QASSCF-POVM (blue dash-dotted lines) together with the estimated perturbative energy correction PT2-POVM (cyan dash-dotted lines). As can be seen, our IC-data based NEVPT2 approach is able to reach a fairly good agreement compared to the classical reference calculations: CASSCF (red solid lines) and PT2 (brown solid lines). Moreover, the PT2 PECs are not only close to the FCI one (black solid lines) but also exhibit -- on a visual scale -- a minimal non-parallelity error. The latter is a strong indication that our SC-NEVPT2 energy corrections include the dominating dynamical electron correlation contributions for LiH along the considered section of the PEC. In addition to the PEC, we also report in Figures \ref{fig:lih-dilation} and \ref{fig:lih-restricted}, respectively, the total numbers of shots required (in 1000s) for the POVM measurement to estimate the ground state energy at given threshold error $\sigma\approx 1.6 \times 10^{-3}$  and  $\sigma\approx 0.8 \times 10^{-3}$, left and right panel, respectively. As expected, the 4P-POVM performs slightly better than the 8P-POVM by demanding for all the bond-distances considered approximately half of the number of shots to reach the same accuracy in the evaluation. To gain a more quantitative insight into the potential and applicability of our method, we report in the bottom panel of Fig.~\ref{fig:lih-dilation} and ~\ref{fig:lih-restricted} the absolute energy difference between the classical reference calculations from PySCF and the POVM estimations for the converged energy (green) and the perturbative energy correction (black) together with the relative threshold error (red). For the energy estimation, we find that both parametrizations of the POVM exhibit  deviations from the reference calculations that are stable and consistent along the PEC. Notably, their magnitudes fluctuate around the preset  threshold error. The latter quantity, assuming that the statistics in the estimations are  approximately normally distributed, is nothing but one standard deviation from the mean of the distribution, thus showing us only a $68\%$ confidence level and resulting in such fluctuations for some of the inter-nuclear distances considered.  Remarkably, the error in the estimation of the energy correction follows a similar trend as the one of the ground state energy in the CAS. Indeed, as the computation of higher order RDMs and consequently of the perturbed energy is carried out in post-processing using a POVM optimized for another observable, reaching a similar value in the precision of the estimation is particularly non trivial. Furthermore, achieving such precision with a very small ratio between the number of shots reached by the POVM optimization to estimate the ground state energy at given threshold error over the total number of
4-RDM matrix elements in the active space $S/n_a^8$ is one of the main results for the proposed method. As an example, if we look at $\ch{LiH}$ at bond length $R_{\rm Li-H} = 1.60 \text{\AA}$ and a fixed threshold error $\sigma_{th} = 0.8 \rm mHa $, the ratio for the two classes considered 8P-POVM and 4P-POVM is $S/n_a^8 \approx 0.15$ and $S/n_a^8 \approx 0.10$ respectively.

\begin{figure}[H]
  \subfigure[
  8P-POVM measurement estimation of the ground state energy with requested threshold error $\sigma = 1.6\times 10^{-3}\,\mathrm{Ha}$.\label{fig:n2-dilation}]{%
    \includegraphics[width=0.45\textwidth]{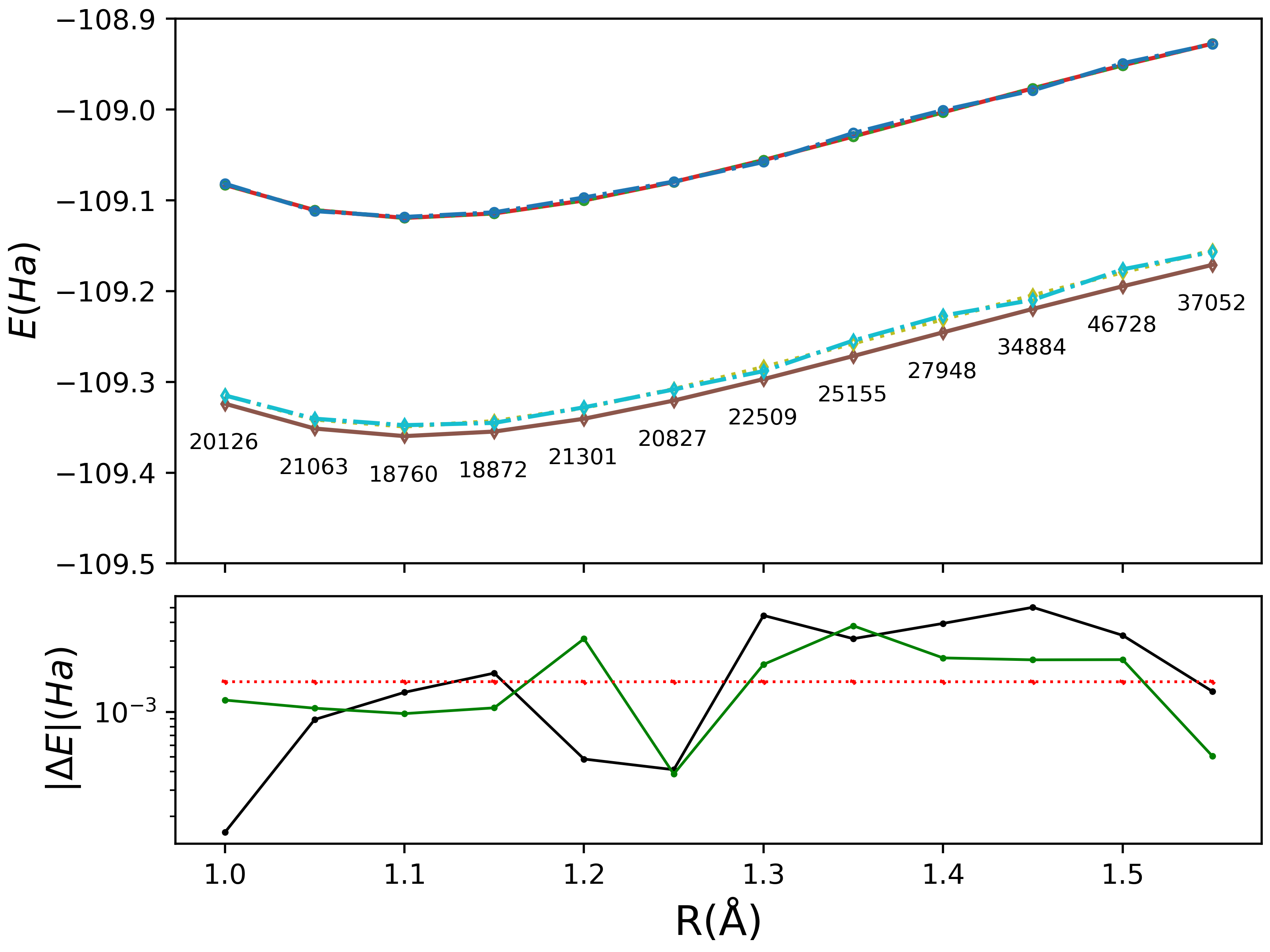}
  }
  \hfill
  \subfigure[
  4P-POVM measurement estimation of the ground state energy with requested threshold error $\sigma = 1.6\times 10^{-3}\,\mathrm{Ha}$.\label{fig:n2-restricted}]{%
    \includegraphics[width=0.45\textwidth]{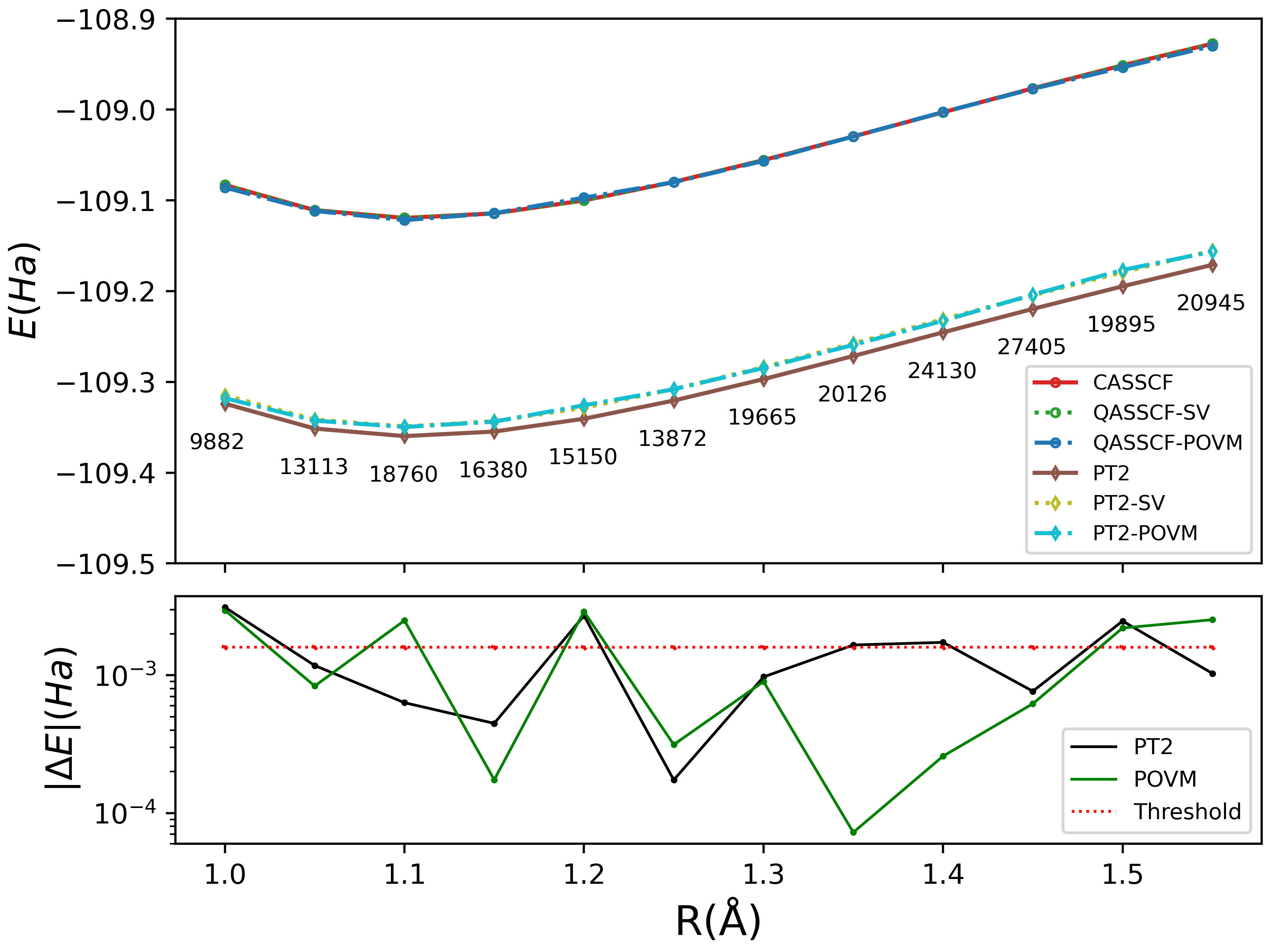}
  }
  \caption{Potential energy curves for \ch{N2} \emph{via} optimized POVMs for requested threshold error $\sigma = 1.6\times 10^{-3}\,\mathrm{Ha}$. Left panel, 8-parameter POVM, right panel, 4-parameter POVM.
  \textbf{Top} Converged noiseless simulation QASSCF-SV (green, dotted) and shots-noise estimation QASSCF-POVM (blue, dash-dotted) together with the estimated perturbative energy correction PT2-POVM (cyan, dash-dotted) compared with the classical reference CASSCF (red, solid), PT2 (brown, solid) and FCI (black, solid) calculations. The numbers below the FCI curves show the total amount of shots required (in 1000's) for the POVM measurement to estimate the ground state energy at given threshold error. \textbf{Bottom} Absolute energy difference between the classical calculation and the POVM estimation for the converged energy (green) and the perturbative energy correction (black) together with the relative threshold error (red).}
  \label{fig:n2}
\end{figure}

As a second molecular example, we consider the ground state PES of $\ch{N_2}$. We choose a cc-pVTZ basis set \cite{dunn89} and consider an active space of six electrons in six orbitals (CAS$(6e,6o)$), correlating the valence $\sigma$\ and $\pi$ framework of $\ch{N_2}$. For this particular case, we sample the perturbative energy correction with the two parametrizations of the POVM considered in the previous example, but we only study the threshold error set to $\sigma\approx 1.6 \times 10^{-3}$ for a resource efficient implementation. 
Figure~\ref{fig:n2-restricted} shows the second order perturbative energy correction for $\ch{N_2}$ as obtained from our quantum-centric NEVPT2 implementation with the input higher-order RDMs estimated in post-processing via an optimized 4-parameter POVM. The top panel in Figure ~\ref{fig:n2-restricted} highlights the PEC as obtained from converged noiseless simulations within the QASSCF-SV scheme (green dotted lines) as well as shot-noise estimations via POVM sampling QASSCF-POVM (blue dash-dotted lines) together with the estimated perturbative energy correction IC-POVM NEVPT2 (cyan dash-dotted lines). We find that our  proposed computational scheme is able to  reach a good accuracy compared to the classical reference calculations: CASSCF (red solid lines) and CASSCF-NEVPT2 (brown solid lines). Likewise, as in the previous example, the numbers below the curve indicate the total number of shots required for the POVM measurement to estimate the ground state energy at the given threshold error (in 1000s). The bottom panel in Figure~\ref{fig:n2-restricted} illustrates the absolute energy difference between the classical reference calculation and the POVM estimation for the converged energy (green) and the perturbative energy correction (black) together with the relative threshold error (red). Similarly to what we observed before, the deviations in the energy estimation from the reference values, as well as the differences in the evaluation of the energy corrections compared to their classical counterpart, are proportional to the fixed threshold error. Furthermore, we are able to achieve such precision with the ratio between the number of shots over the number of matrix elements estimated of $S/n_a^8 \approx 0.041$ and $S/n_a^8 \approx 0.033$  for the 8P and 4P POVMs respectively, confirming the robustness of the method for systems with active spaces up to 12 spin-orbitals. 
\onecolumngrid

\twocolumngrid    

\vspace{0.2cm}

\subsection{Scaling with molecule size: \ch{H}-chains and polyenes}\label{sec:results:polyenes}

To study the scaling of the proposed algorithm with the overall system size, we consider in this section polyenes of varying system size, a class of molecules with alternating single- and double-bonds \ch{C_{2n} H_{2n+2}} for n up to 3, as well as linear hydrogen chains \ch{H_n} for n up to 7. These are benchmark systems with 8, 10, 12, and 14 qubits respectively. Results for the butadiene molecule (\ch{C_{4} H_{6}}) are presented in Table~\ref{tab:butadiene} and for hexatriene ($\ch{C_{6} H_{8}}$) in Table \ref{tab:hexatriene}. The tables summarize the ground state energy estimated via POVMs ($E_{VQESCF}$), the PT2 energy estimated via POVMs ($E_{POVM/PT2}$), and the errors in these estimations with respect to the classically calculated values, along with the number of measurement shots needed. For these systems, we note that the ratio, provided the 8P-POVM is optimised for a standard error of $\sigma_{th} = 1.6 \rm mHa $, is $S/n_a^8 \approx 0.14$ and $S/n_a^8 \approx 0.02$ for the 8 and 12-qubit examples respectively. By contrast, a restricted parametrization 4P-POVM yields a ratio of $S/n_a^8 \approx 0.08$ and $S/n_a^8 \approx 0.01$ for the two system-sizes considered in this work. 

\onecolumngrid

\begin{table}[H]
\caption{Ground-state energies for butadiene in a CAS($4e,4o$) with a cc-pVTZ basis for C and H. Classical reference results have been obtained with PySCF: $E_{\mathrm{CASSCF}} = -155.03106\,\mathrm{Ha}$ and $E_{\mathrm{PT2}} = -0.66324\,\mathrm{Ha}$.}
\label{tab:butadiene}
\centering
\begin{ruledtabular}
\begin{tabular}{c l c c c c c c}
    POVM Class               & $\sigma_{\mathrm{th}}$ (mHa) & $E_{VQESCF}$ (Ha) & $E_{PT2}^{POVM}$  (Ha) &   Shots ($\times 10^3$) & $\Delta E$ (mHa) & $\Delta E_{PT2}$ (mHa) \\
\colrule
    \multirow{2}{*}{8P-POVM} & 1.60 & -155.08775 & -0.65749 & 2380 & 0.30  & 5.88\\
                             & 3.16 & -155.08964 & -0.65628  & 630 & 2.19 & 7.09\\
    \hline
    \multirow{2}{*}{4P-POVM} & 1.60 & -155.08555 & -0.66479 & 1335 & 1.89 & 1.42\\
                             & 3.16 & -155.08981 & -0.67579 & 315 & 2.36  & 12.42 \\
\end{tabular}
\end{ruledtabular}
\end{table}

\begin{table}[H]
\caption{Ground-state energies for hexatriene in a CAS($6e,6o$) with a cc-pVTZ basis for C and H. Classical reference results have been obtained with PySCF: $E_{\mathrm{CASSCF}} = -231.97154\,\mathrm{Ha}$ and $E_{\mathrm{PT2}} = -0.98576\,\mathrm{Ha}$.}
\label{tab:hexatriene}
\centering
\begin{ruledtabular}
\begin{tabular}{c l c c c c c c}
    POVM Class               & $\sigma_{\mathrm{th}}$ (mHa) & $E_{VQESCF}$ (Ha) & $E_{PT2}^{POVM}$  (Ha) &   Shots ($\times 10^3$) & $\Delta E$ (mHa) & $\Delta E_{PT2}$ (mHa) \\
\colrule
    \multirow{2}{*}{8P-POVM} & 1.60 & -232.05634  & -0.99024 & 9087 & 0.20 & 4.27\\
                             & 3.16 & -232.05330  & -0.99542 & 3015 & 2.84 & 9.45\\
    \hline
    \multirow{2}{*}{4P-POVM} & 1.60 & -232.05412 & -0.98424 & 6048 & 2.02 & 1.73\\
                             & 3.16 & -232.05647 & -0.99289 & 1890 & 0.33  & 6.92 \\
\end{tabular}
\end{ruledtabular}
\end{table}

\begin{figure}[H]
    \centering
    \includegraphics[width = 0.60\textwidth]{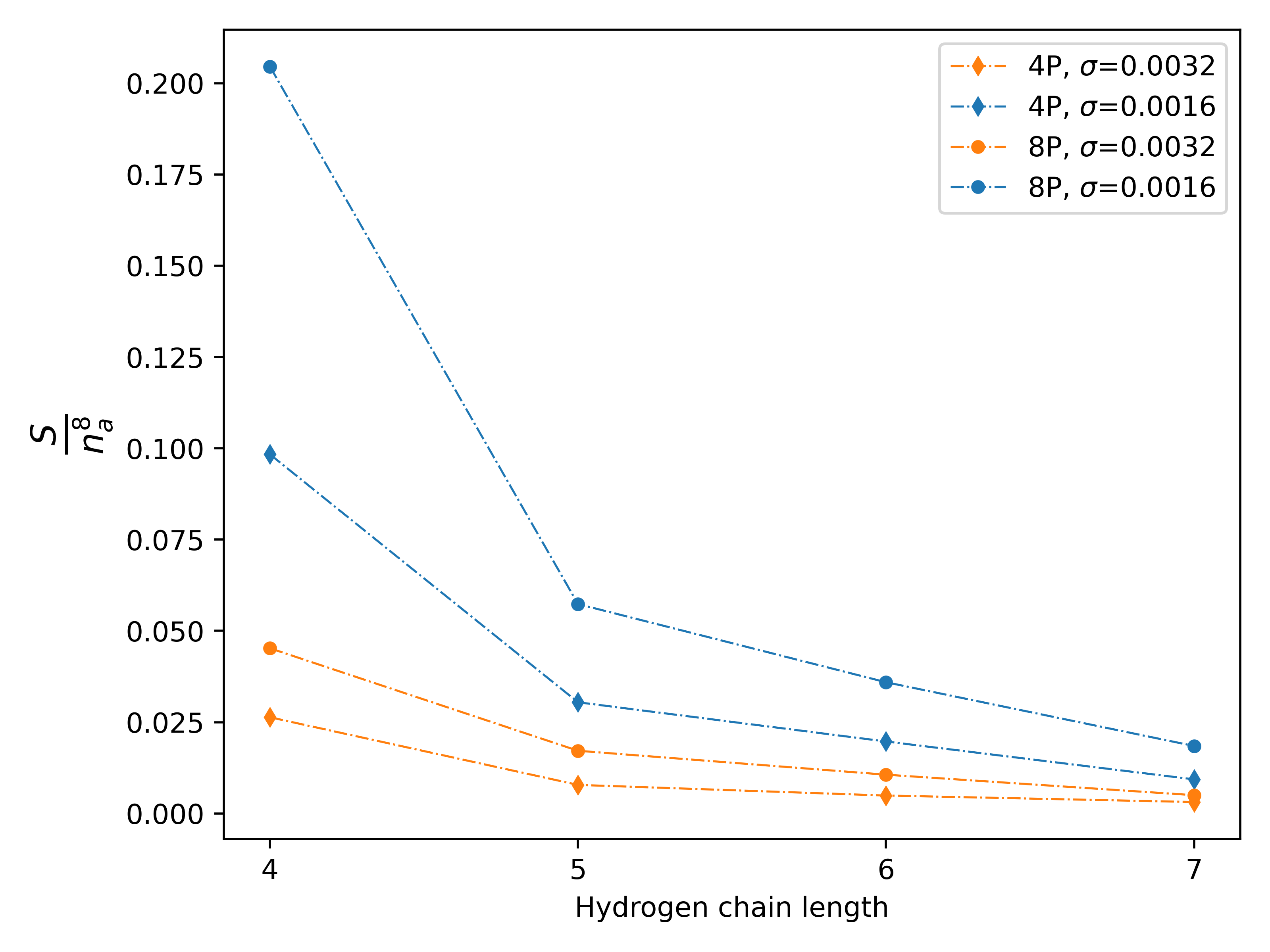}
    \caption{Ratio of the shots required to estimate the ground state energy for given standard-error to the number of 4-RDM matrix elements estimated for increasing length of linear hydrogen chains. The 8P-POVM parametrization takes roughly double the amount of shots to reach a given standard error compare to the 4P-POVM one, while the ratio becomes more favourable with increasing scale in all cases. This data shows the quantum measurement cost for such calculations is greatly mitigated by the usage of IC-POVMs. 
    }
    \label{fig:scaling}
\end{figure}

\twocolumngrid

In contrast to the diatomic examples discussed in Section \ref{sec:results:diatomics}, we find that in case of the polyenes the ratio of $S/n^{8}_a$ becomes even more favorable as the system size increases. To provide a more comprehensive understanding of this advantageous scaling, we extended our calculations to assess the same ratio for linear hydrogen chains of increasing length. Figure~\ref{fig:scaling} genuinely illustrates this trend, indicating a swift enhancement in the ratio as the system size grows. These findings therefore demonstrate the high and favourable efficiency of the scheme, offering a significant reduction in the measurement overhead, especially for near-term quantum devices.

\section{Conclusions}
In this article, we introduced and assessed a quantum-centric computational approach to efficiently grasp both static and dynamic electron correlation effects in molecular compounds by means of an accurate estimation of perturbative energy corrections subsequent to an active-space driven quantum correlation approach. The robustness of our proposed approach was examined on the basis of comparing our quantum simulations with classical CASSCF+SC-NEVPT2 calculations, underlining the potential of our approach to offer insights within a real-world scientific framework.
The efficiency of our approach is particularly evident when considering the effects of shot-noise, an intrinsic limitation of any quantum-centric approach to calculate molecular energies. Moreover, we demonstrated the versatility of our method not only with simple diatomic molecules but also considered extended molecular compounds such as polyenes which are surrogate models for bioactive compounds such as carotenoids, further emphasizing the universal applicability of our computational approach to molecular systems with a genuine complex (ground-state) electronic structure. Among the measurement parametrizations tested, certain versions showcased exceptional resource efficiency, indicating a promising direction for future efforts.
We emphasized in this work the accuracy of informationally-complete measurement approach to compute higher-order reduced density matrices despite the applied measurement strategy being optimized for a reduced variance in the estimation of the ground state energy solely dominated by low-order reduced density matrices. Notably, our measurement scheme could also be optimized for a reduced variance in a different operator of choice, such as the number operator, which could potentially lead to better estimations of (higher-order) reduced density matrix elements. 
A striking observation was the improved efficiency witnessed with larger molecular systems. This favorable scaling suggests that as one moves to more complex quantum systems, the proposed strategy could present even more pronounced benefits, potentially offering significant computational savings in a quantum-centric HPC-driven post-processing setting. Further, our computational model does not require any extra quantum resources. The IC-POVM measurement outcomes obtained in the estimation of the ground state energy can  simply be recycled to compute the perturbative correction, with the required three- and four-body reduced density matrix elements easily estimated in an embarrassingly parallel fashion. One can also see how the estimation of the RDM elements can also be used at each ADAPT-VQE iteration to not only add new gates but calculate optimal orbital rotations, allowing for efficient implementation of the VQE-SCF routine outlined in \cite{fitzpatrick2022selfconsistent}. 
Additionally, our findings unequivocally underscore the potential of our model to be readily employed in quantum chemistry experiments on near-term quantum devices, a possibility we are currently pursuing in our ongoing efforts on accurate calculations of excitation energies on quantum computers for molecular systems with application in photodynamic therapy \cite{pavo24b}. By maintaining a high precision with a minimal measurement overhead, our approach constitutes a crucial advancement, especially given the inherent limited capabilities of present-day quantum hardware.
Finally, our quantum-centric model combining the best of both, quantum and traditional  computational techniques in quantum chemistry emphasizes a promising trajectory for the future of quantum chemistry and molecular simulations. Given the scalable and efficient nature of our proposed method, an in-depth exploration into the chemistry of more more complex molecular systems is warranted, potentially paving the way for breakthroughs in quantum computation.

\section*{ACKNOWLEDGEMENTS}
The authors thank Guillermo Garc{\'i}a-P{\'e}rez, Alessandro Lunghi and Keijo Korhonen for useful discussion and insights to improve this work. Work on ''Quantum Computing for Photon-Drug Interactions in Cancer Prevention and
Treatment" is supported by Wellcome Leap as part of the Q4Bio Program.

\section*{Author contributions}
A.~F.~and N.~W.~T.~proposed the methods, wrote the code and the first draft. R.~.D.~R.~E.~generalised the code for arbitrary-order reduced density matrices. A.~F. and N.~W.~T. prepared the
figures. A.~F. and N.~W.~T. obtained, evaluated, and analyzed the data. All authors participated
in the discussion and writing of the manuscript.

\bibliography{bibliography}

\begin{thebibliography}{60}%
\makeatletter
\providecommand \@ifxundefined [1]{%
 \@ifx{#1\undefined}
}%
\providecommand \@ifnum [1]{%
 \ifnum #1\expandafter \@firstoftwo
 \else \expandafter \@secondoftwo
 \fi
}%
\providecommand \@ifx [1]{%
 \ifx #1\expandafter \@firstoftwo
 \else \expandafter \@secondoftwo
 \fi
}%
\providecommand \natexlab [1]{#1}%
\providecommand \enquote  [1]{``#1''}%
\providecommand \bibnamefont  [1]{#1}%
\providecommand \bibfnamefont [1]{#1}%
\providecommand \citenamefont [1]{#1}%
\providecommand \href@noop [0]{\@secondoftwo}%
\providecommand \href [0]{\begingroup \@sanitize@url \@href}%
\providecommand \@href[1]{\@@startlink{#1}\@@href}%
\providecommand \@@href[1]{\endgroup#1\@@endlink}%
\providecommand \@sanitize@url [0]{\catcode `\\12\catcode `\$12\catcode
  `\&12\catcode `\#12\catcode `\^12\catcode `\_12\catcode `\%12\relax}%
\providecommand \@@startlink[1]{}%
\providecommand \@@endlink[0]{}%
\providecommand \url  [0]{\begingroup\@sanitize@url \@url }%
\providecommand \@url [1]{\endgroup\@href {#1}{\urlprefix }}%
\providecommand \urlprefix  [0]{URL }%
\providecommand \Eprint [0]{\href }%
\providecommand \doibase [0]{http://dx.doi.org/}%
\providecommand \selectlanguage [0]{\@gobble}%
\providecommand \bibinfo  [0]{\@secondoftwo}%
\providecommand \bibfield  [0]{\@secondoftwo}%
\providecommand \translation [1]{[#1]}%
\providecommand \BibitemOpen [0]{}%
\providecommand \bibitemStop [0]{}%
\providecommand \bibitemNoStop [0]{.\EOS\space}%
\providecommand \EOS [0]{\spacefactor3000\relax}%
\providecommand \BibitemShut  [1]{\csname bibitem#1\endcsname}%
\let\auto@bib@innerbib\@empty
%</preamble>
\bibitem [{\citenamefont {Lanyon}\ \emph {et~al.}(2010)\citenamefont {Lanyon},
  \citenamefont {Whitfield}, \citenamefont {Gillett}, \citenamefont {Goggin},
  \citenamefont {Almeida}, \citenamefont {Kassal}, \citenamefont {Biamonte},
  \citenamefont {Mohseni}, \citenamefont {Powell}, \citenamefont {Barbieri},
  \citenamefont {Aspuru-Guzik},\ and\ \citenamefont {White}}]{Lanyon2010}%
  \BibitemOpen
  \bibfield  {author} {\bibinfo {author} {\bibfnamefont {B.~P.}\ \bibnamefont
  {Lanyon}}, \bibinfo {author} {\bibfnamefont {J.~D.}\ \bibnamefont
  {Whitfield}}, \bibinfo {author} {\bibfnamefont {G.~G.}\ \bibnamefont
  {Gillett}}, \bibinfo {author} {\bibfnamefont {M.~E.}\ \bibnamefont {Goggin}},
  \bibinfo {author} {\bibfnamefont {M.~P.}\ \bibnamefont {Almeida}}, \bibinfo
  {author} {\bibfnamefont {I.}~\bibnamefont {Kassal}}, \bibinfo {author}
  {\bibfnamefont {J.~D.}\ \bibnamefont {Biamonte}}, \bibinfo {author}
  {\bibfnamefont {M.}~\bibnamefont {Mohseni}}, \bibinfo {author} {\bibfnamefont
  {B.~J.}\ \bibnamefont {Powell}}, \bibinfo {author} {\bibfnamefont
  {M.}~\bibnamefont {Barbieri}}, \bibinfo {author} {\bibfnamefont
  {A.}~\bibnamefont {Aspuru-Guzik}}, \ and\ \bibinfo {author} {\bibfnamefont
  {A.~G.}\ \bibnamefont {White}},\ }\href {\doibase 10.1038/nchem.483}
  {\bibfield  {journal} {\bibinfo  {journal} {Nat. Chem.}\ }\textbf {\bibinfo
  {volume} {2}},\ \bibinfo {pages} {106–111} (\bibinfo {year}
  {2010})}\BibitemShut {NoStop}%
\bibitem [{\citenamefont {Peruzzo}\ \emph {et~al.}(2014)\citenamefont
  {Peruzzo}, \citenamefont {McClean}, \citenamefont {Shadbolt}, \citenamefont
  {Yung}, \citenamefont {Zhou}, \citenamefont {Love}, \citenamefont
  {Aspuru-Guzik},\ and\ \citenamefont {O’Brien}}]{peru14a}%
  \BibitemOpen
  \bibfield  {author} {\bibinfo {author} {\bibfnamefont {A.}~\bibnamefont
  {Peruzzo}}, \bibinfo {author} {\bibfnamefont {J.}~\bibnamefont {McClean}},
  \bibinfo {author} {\bibfnamefont {P.}~\bibnamefont {Shadbolt}}, \bibinfo
  {author} {\bibfnamefont {M.-H.}\ \bibnamefont {Yung}}, \bibinfo {author}
  {\bibfnamefont {X.-Q.}\ \bibnamefont {Zhou}}, \bibinfo {author}
  {\bibfnamefont {P.~J.}\ \bibnamefont {Love}}, \bibinfo {author}
  {\bibfnamefont {A.}~\bibnamefont {Aspuru-Guzik}}, \ and\ \bibinfo {author}
  {\bibfnamefont {J.~L.}\ \bibnamefont {O’Brien}},\ }\href {\doibase
  10.1038/ncomms5213} {\bibfield  {journal} {\bibinfo  {journal}
  {Nat.~Commun.}\ }\textbf {\bibinfo {volume} {5}},\ \bibinfo {pages} {1}
  (\bibinfo {year} {2014})}\BibitemShut {NoStop}%
\bibitem [{\citenamefont {Kandala}\ \emph {et~al.}(2017)\citenamefont
  {Kandala}, \citenamefont {Mezzacapo}, \citenamefont {Temme}, \citenamefont
  {Takita}, \citenamefont {Brink}, \citenamefont {Chow},\ and\ \citenamefont
  {Gambetta}}]{Kandala2017}%
  \BibitemOpen
  \bibfield  {author} {\bibinfo {author} {\bibfnamefont {A.}~\bibnamefont
  {Kandala}}, \bibinfo {author} {\bibfnamefont {A.}~\bibnamefont {Mezzacapo}},
  \bibinfo {author} {\bibfnamefont {K.}~\bibnamefont {Temme}}, \bibinfo
  {author} {\bibfnamefont {M.}~\bibnamefont {Takita}}, \bibinfo {author}
  {\bibfnamefont {M.}~\bibnamefont {Brink}}, \bibinfo {author} {\bibfnamefont
  {J.~M.}\ \bibnamefont {Chow}}, \ and\ \bibinfo {author} {\bibfnamefont
  {J.~M.}\ \bibnamefont {Gambetta}},\ }\href {\doibase 10.1038/nature23879}
  {\bibfield  {journal} {\bibinfo  {journal} {Nature}\ }\textbf {\bibinfo
  {volume} {549}},\ \bibinfo {pages} {242} (\bibinfo {year}
  {2017})}\BibitemShut {NoStop}%
\bibitem [{\citenamefont {Scheurer}\ \emph {et~al.}(2023)\citenamefont
  {Scheurer}, \citenamefont {Anselmetti}, \citenamefont {Oumarou},
  \citenamefont {Gogolin},\ and\ \citenamefont {Rubin}}]{sche23c}%
  \BibitemOpen
  \bibfield  {author} {\bibinfo {author} {\bibfnamefont {M.}~\bibnamefont
  {Scheurer}}, \bibinfo {author} {\bibfnamefont {G.-L.~R.}\ \bibnamefont
  {Anselmetti}}, \bibinfo {author} {\bibfnamefont {O.}~\bibnamefont {Oumarou}},
  \bibinfo {author} {\bibfnamefont {C.}~\bibnamefont {Gogolin}}, \ and\
  \bibinfo {author} {\bibfnamefont {N.~C.}\ \bibnamefont {Rubin}},\ }\href
  {\doibase 10.48550/ARXIV.2312.08110} {\enquote {\bibinfo {title} {Tailored
  and externally corrected coupled cluster with quantum inputs},}\ } (\bibinfo
  {year} {2023})\BibitemShut {NoStop}%
\bibitem [{\citenamefont {O’Brien}\ \emph {et~al.}(2023)\citenamefont
  {O’Brien}, \citenamefont {Anselmetti}, \citenamefont {Gkritsis},
  \citenamefont {Elfving}, \citenamefont {Polla}, \citenamefont {Huggins},
  \citenamefont {Oumarou}, \citenamefont {Kechedzhi}, \citenamefont {Abanin},
  \citenamefont {Acharya}, \citenamefont {Aleiner}, \citenamefont {Allen},
  \citenamefont {Andersen}, \citenamefont {Anderson}, \citenamefont {Ansmann},
  \citenamefont {Arute}, \citenamefont {Arya}, \citenamefont {Asfaw},
  \citenamefont {Atalaya}, \citenamefont {Bardin}, \citenamefont {Bengtsson},
  \citenamefont {Bortoli}, \citenamefont {Bourassa}, \citenamefont {Bovaird},
  \citenamefont {Brill}, \citenamefont {Broughton}, \citenamefont {Buckley},
  \citenamefont {Buell}, \citenamefont {Burger}, \citenamefont {Burkett},
  \citenamefont {Bushnell}, \citenamefont {Campero}, \citenamefont {Chen},
  \citenamefont {Chiaro}, \citenamefont {Chik}, \citenamefont {Cogan},
  \citenamefont {Collins}, \citenamefont {Conner}, \citenamefont {Courtney},
  \citenamefont {Crook}, \citenamefont {Curtin}, \citenamefont {Debroy},
  \citenamefont {Demura}, \citenamefont {Drozdov}, \citenamefont {Dunsworth},
  \citenamefont {Erickson}, \citenamefont {Faoro}, \citenamefont {Farhi},
  \citenamefont {Fatemi}, \citenamefont {Ferreira}, \citenamefont
  {Flores~Burgos}, \citenamefont {Forati}, \citenamefont {Fowler},
  \citenamefont {Foxen}, \citenamefont {Giang}, \citenamefont {Gidney},
  \citenamefont {Gilboa}, \citenamefont {Giustina}, \citenamefont {Gosula},
  \citenamefont {Grajales~Dau}, \citenamefont {Gross}, \citenamefont
  {Habegger}, \citenamefont {Hamilton}, \citenamefont {Hansen}, \citenamefont
  {Harrigan}, \citenamefont {Harrington}, \citenamefont {Heu}, \citenamefont
  {Hoffmann}, \citenamefont {Hong}, \citenamefont {Huang}, \citenamefont
  {Huff}, \citenamefont {Ioffe}, \citenamefont {Isakov}, \citenamefont
  {Iveland}, \citenamefont {Jeffrey}, \citenamefont {Jiang}, \citenamefont
  {Jones}, \citenamefont {Juhas}, \citenamefont {Kafri}, \citenamefont
  {Khattar}, \citenamefont {Khezri}, \citenamefont {Kieferová}, \citenamefont
  {Kim}, \citenamefont {Klimov}, \citenamefont {Klots}, \citenamefont
  {Korotkov}, \citenamefont {Kostritsa}, \citenamefont {Kreikebaum},
  \citenamefont {Landhuis}, \citenamefont {Laptev}, \citenamefont {Lau},
  \citenamefont {Laws}, \citenamefont {Lee}, \citenamefont {Lee}, \citenamefont
  {Lester}, \citenamefont {Lill}, \citenamefont {Liu}, \citenamefont
  {Livingston}, \citenamefont {Locharla}, \citenamefont {Malone}, \citenamefont
  {Mandrà}, \citenamefont {Martin}, \citenamefont {Martin}, \citenamefont
  {McClean}, \citenamefont {McCourt}, \citenamefont {McEwen}, \citenamefont
  {Mi}, \citenamefont {Mieszala}, \citenamefont {Miao}, \citenamefont
  {Mohseni}, \citenamefont {Montazeri}, \citenamefont {Morvan}, \citenamefont
  {Movassagh}, \citenamefont {Mruczkiewicz}, \citenamefont {Naaman},
  \citenamefont {Neeley}, \citenamefont {Neill}, \citenamefont {Nersisyan},
  \citenamefont {Newman}, \citenamefont {Ng}, \citenamefont {Nguyen},
  \citenamefont {Nguyen}, \citenamefont {Niu}, \citenamefont {Omonije},
  \citenamefont {Opremcak}, \citenamefont {Petukhov}, \citenamefont {Potter},
  \citenamefont {Pryadko}, \citenamefont {Quintana}, \citenamefont {Rocque},
  \citenamefont {Roushan}, \citenamefont {Saei}, \citenamefont {Sank},
  \citenamefont {Sankaragomathi}, \citenamefont {Satzinger}, \citenamefont
  {Schurkus}, \citenamefont {Schuster}, \citenamefont {Shearn}, \citenamefont
  {Shorter}, \citenamefont {Shutty}, \citenamefont {Shvarts}, \citenamefont
  {Skruzny}, \citenamefont {Smith}, \citenamefont {Somma}, \citenamefont
  {Sterling}, \citenamefont {Strain}, \citenamefont {Szalay}, \citenamefont
  {Thor}, \citenamefont {Torres}, \citenamefont {Vidal}, \citenamefont
  {Villalonga}, \citenamefont {Vollgraff~Heidweiller}, \citenamefont {White},
  \citenamefont {Woo}, \citenamefont {Xing}, \citenamefont {Yao}, \citenamefont
  {Yeh}, \citenamefont {Yoo}, \citenamefont {Young}, \citenamefont {Zalcman},
  \citenamefont {Zhang}, \citenamefont {Zhu}, \citenamefont {Zobrist},
  \citenamefont {Bacon}, \citenamefont {Boixo}, \citenamefont {Chen},
  \citenamefont {Hilton}, \citenamefont {Kelly}, \citenamefont {Lucero},
  \citenamefont {Megrant}, \citenamefont {Neven}, \citenamefont {Smelyanskiy},
  \citenamefont {Gogolin}, \citenamefont {Babbush},\ and\ \citenamefont
  {Rubin}}]{OBrien2023}%
  \BibitemOpen
  \bibfield  {author} {\bibinfo {author} {\bibfnamefont {T.~E.}\ \bibnamefont
  {O’Brien}}, \bibinfo {author} {\bibfnamefont {G.}~\bibnamefont
  {Anselmetti}}, \bibinfo {author} {\bibfnamefont {F.}~\bibnamefont
  {Gkritsis}}, \bibinfo {author} {\bibfnamefont {V.~E.}\ \bibnamefont
  {Elfving}}, \bibinfo {author} {\bibfnamefont {S.}~\bibnamefont {Polla}},
  \bibinfo {author} {\bibfnamefont {W.~J.}\ \bibnamefont {Huggins}}, \bibinfo
  {author} {\bibfnamefont {O.}~\bibnamefont {Oumarou}}, \bibinfo {author}
  {\bibfnamefont {K.}~\bibnamefont {Kechedzhi}}, \bibinfo {author}
  {\bibfnamefont {D.}~\bibnamefont {Abanin}}, \bibinfo {author} {\bibfnamefont
  {R.}~\bibnamefont {Acharya}}, \bibinfo {author} {\bibfnamefont
  {I.}~\bibnamefont {Aleiner}}, \bibinfo {author} {\bibfnamefont
  {R.}~\bibnamefont {Allen}}, \bibinfo {author} {\bibfnamefont {T.~I.}\
  \bibnamefont {Andersen}}, \bibinfo {author} {\bibfnamefont {K.}~\bibnamefont
  {Anderson}}, \bibinfo {author} {\bibfnamefont {M.}~\bibnamefont {Ansmann}},
  \bibinfo {author} {\bibfnamefont {F.}~\bibnamefont {Arute}}, \bibinfo
  {author} {\bibfnamefont {K.}~\bibnamefont {Arya}}, \bibinfo {author}
  {\bibfnamefont {A.}~\bibnamefont {Asfaw}}, \bibinfo {author} {\bibfnamefont
  {J.}~\bibnamefont {Atalaya}}, \bibinfo {author} {\bibfnamefont {J.~C.}\
  \bibnamefont {Bardin}}, \bibinfo {author} {\bibfnamefont {A.}~\bibnamefont
  {Bengtsson}}, \bibinfo {author} {\bibfnamefont {G.}~\bibnamefont {Bortoli}},
  \bibinfo {author} {\bibfnamefont {A.}~\bibnamefont {Bourassa}}, \bibinfo
  {author} {\bibfnamefont {J.}~\bibnamefont {Bovaird}}, \bibinfo {author}
  {\bibfnamefont {L.}~\bibnamefont {Brill}}, \bibinfo {author} {\bibfnamefont
  {M.}~\bibnamefont {Broughton}}, \bibinfo {author} {\bibfnamefont
  {B.}~\bibnamefont {Buckley}}, \bibinfo {author} {\bibfnamefont {D.~A.}\
  \bibnamefont {Buell}}, \bibinfo {author} {\bibfnamefont {T.}~\bibnamefont
  {Burger}}, \bibinfo {author} {\bibfnamefont {B.}~\bibnamefont {Burkett}},
  \bibinfo {author} {\bibfnamefont {N.}~\bibnamefont {Bushnell}}, \bibinfo
  {author} {\bibfnamefont {J.}~\bibnamefont {Campero}}, \bibinfo {author}
  {\bibfnamefont {Z.}~\bibnamefont {Chen}}, \bibinfo {author} {\bibfnamefont
  {B.}~\bibnamefont {Chiaro}}, \bibinfo {author} {\bibfnamefont
  {D.}~\bibnamefont {Chik}}, \bibinfo {author} {\bibfnamefont {J.}~\bibnamefont
  {Cogan}}, \bibinfo {author} {\bibfnamefont {R.}~\bibnamefont {Collins}},
  \bibinfo {author} {\bibfnamefont {P.}~\bibnamefont {Conner}}, \bibinfo
  {author} {\bibfnamefont {W.}~\bibnamefont {Courtney}}, \bibinfo {author}
  {\bibfnamefont {A.~L.}\ \bibnamefont {Crook}}, \bibinfo {author}
  {\bibfnamefont {B.}~\bibnamefont {Curtin}}, \bibinfo {author} {\bibfnamefont
  {D.~M.}\ \bibnamefont {Debroy}}, \bibinfo {author} {\bibfnamefont
  {S.}~\bibnamefont {Demura}}, \bibinfo {author} {\bibfnamefont
  {I.}~\bibnamefont {Drozdov}}, \bibinfo {author} {\bibfnamefont
  {A.}~\bibnamefont {Dunsworth}}, \bibinfo {author} {\bibfnamefont
  {C.}~\bibnamefont {Erickson}}, \bibinfo {author} {\bibfnamefont
  {L.}~\bibnamefont {Faoro}}, \bibinfo {author} {\bibfnamefont
  {E.}~\bibnamefont {Farhi}}, \bibinfo {author} {\bibfnamefont
  {R.}~\bibnamefont {Fatemi}}, \bibinfo {author} {\bibfnamefont {V.~S.}\
  \bibnamefont {Ferreira}}, \bibinfo {author} {\bibfnamefont {L.}~\bibnamefont
  {Flores~Burgos}}, \bibinfo {author} {\bibfnamefont {E.}~\bibnamefont
  {Forati}}, \bibinfo {author} {\bibfnamefont {A.~G.}\ \bibnamefont {Fowler}},
  \bibinfo {author} {\bibfnamefont {B.}~\bibnamefont {Foxen}}, \bibinfo
  {author} {\bibfnamefont {W.}~\bibnamefont {Giang}}, \bibinfo {author}
  {\bibfnamefont {C.}~\bibnamefont {Gidney}}, \bibinfo {author} {\bibfnamefont
  {D.}~\bibnamefont {Gilboa}}, \bibinfo {author} {\bibfnamefont
  {M.}~\bibnamefont {Giustina}}, \bibinfo {author} {\bibfnamefont
  {R.}~\bibnamefont {Gosula}}, \bibinfo {author} {\bibfnamefont
  {A.}~\bibnamefont {Grajales~Dau}}, \bibinfo {author} {\bibfnamefont {J.~A.}\
  \bibnamefont {Gross}}, \bibinfo {author} {\bibfnamefont {S.}~\bibnamefont
  {Habegger}}, \bibinfo {author} {\bibfnamefont {M.~C.}\ \bibnamefont
  {Hamilton}}, \bibinfo {author} {\bibfnamefont {M.}~\bibnamefont {Hansen}},
  \bibinfo {author} {\bibfnamefont {M.~P.}\ \bibnamefont {Harrigan}}, \bibinfo
  {author} {\bibfnamefont {S.~D.}\ \bibnamefont {Harrington}}, \bibinfo
  {author} {\bibfnamefont {P.}~\bibnamefont {Heu}}, \bibinfo {author}
  {\bibfnamefont {M.~R.}\ \bibnamefont {Hoffmann}}, \bibinfo {author}
  {\bibfnamefont {S.}~\bibnamefont {Hong}}, \bibinfo {author} {\bibfnamefont
  {T.}~\bibnamefont {Huang}}, \bibinfo {author} {\bibfnamefont
  {A.}~\bibnamefont {Huff}}, \bibinfo {author} {\bibfnamefont {L.~B.}\
  \bibnamefont {Ioffe}}, \bibinfo {author} {\bibfnamefont {S.~V.}\ \bibnamefont
  {Isakov}}, \bibinfo {author} {\bibfnamefont {J.}~\bibnamefont {Iveland}},
  \bibinfo {author} {\bibfnamefont {E.}~\bibnamefont {Jeffrey}}, \bibinfo
  {author} {\bibfnamefont {Z.}~\bibnamefont {Jiang}}, \bibinfo {author}
  {\bibfnamefont {C.}~\bibnamefont {Jones}}, \bibinfo {author} {\bibfnamefont
  {P.}~\bibnamefont {Juhas}}, \bibinfo {author} {\bibfnamefont
  {D.}~\bibnamefont {Kafri}}, \bibinfo {author} {\bibfnamefont
  {T.}~\bibnamefont {Khattar}}, \bibinfo {author} {\bibfnamefont
  {M.}~\bibnamefont {Khezri}}, \bibinfo {author} {\bibfnamefont
  {M.}~\bibnamefont {Kieferová}}, \bibinfo {author} {\bibfnamefont
  {S.}~\bibnamefont {Kim}}, \bibinfo {author} {\bibfnamefont {P.~V.}\
  \bibnamefont {Klimov}}, \bibinfo {author} {\bibfnamefont {A.~R.}\
  \bibnamefont {Klots}}, \bibinfo {author} {\bibfnamefont {A.~N.}\ \bibnamefont
  {Korotkov}}, \bibinfo {author} {\bibfnamefont {F.}~\bibnamefont {Kostritsa}},
  \bibinfo {author} {\bibfnamefont {J.~M.}\ \bibnamefont {Kreikebaum}},
  \bibinfo {author} {\bibfnamefont {D.}~\bibnamefont {Landhuis}}, \bibinfo
  {author} {\bibfnamefont {P.}~\bibnamefont {Laptev}}, \bibinfo {author}
  {\bibfnamefont {K.-M.}\ \bibnamefont {Lau}}, \bibinfo {author} {\bibfnamefont
  {L.}~\bibnamefont {Laws}}, \bibinfo {author} {\bibfnamefont {J.}~\bibnamefont
  {Lee}}, \bibinfo {author} {\bibfnamefont {K.}~\bibnamefont {Lee}}, \bibinfo
  {author} {\bibfnamefont {B.~J.}\ \bibnamefont {Lester}}, \bibinfo {author}
  {\bibfnamefont {A.~T.}\ \bibnamefont {Lill}}, \bibinfo {author}
  {\bibfnamefont {W.}~\bibnamefont {Liu}}, \bibinfo {author} {\bibfnamefont
  {W.~P.}\ \bibnamefont {Livingston}}, \bibinfo {author} {\bibfnamefont
  {A.}~\bibnamefont {Locharla}}, \bibinfo {author} {\bibfnamefont {F.~D.}\
  \bibnamefont {Malone}}, \bibinfo {author} {\bibfnamefont {S.}~\bibnamefont
  {Mandrà}}, \bibinfo {author} {\bibfnamefont {O.}~\bibnamefont {Martin}},
  \bibinfo {author} {\bibfnamefont {S.}~\bibnamefont {Martin}}, \bibinfo
  {author} {\bibfnamefont {J.~R.}\ \bibnamefont {McClean}}, \bibinfo {author}
  {\bibfnamefont {T.}~\bibnamefont {McCourt}}, \bibinfo {author} {\bibfnamefont
  {M.}~\bibnamefont {McEwen}}, \bibinfo {author} {\bibfnamefont
  {X.}~\bibnamefont {Mi}}, \bibinfo {author} {\bibfnamefont {A.}~\bibnamefont
  {Mieszala}}, \bibinfo {author} {\bibfnamefont {K.~C.}\ \bibnamefont {Miao}},
  \bibinfo {author} {\bibfnamefont {M.}~\bibnamefont {Mohseni}}, \bibinfo
  {author} {\bibfnamefont {S.}~\bibnamefont {Montazeri}}, \bibinfo {author}
  {\bibfnamefont {A.}~\bibnamefont {Morvan}}, \bibinfo {author} {\bibfnamefont
  {R.}~\bibnamefont {Movassagh}}, \bibinfo {author} {\bibfnamefont
  {W.}~\bibnamefont {Mruczkiewicz}}, \bibinfo {author} {\bibfnamefont
  {O.}~\bibnamefont {Naaman}}, \bibinfo {author} {\bibfnamefont
  {M.}~\bibnamefont {Neeley}}, \bibinfo {author} {\bibfnamefont
  {C.}~\bibnamefont {Neill}}, \bibinfo {author} {\bibfnamefont
  {A.}~\bibnamefont {Nersisyan}}, \bibinfo {author} {\bibfnamefont
  {M.}~\bibnamefont {Newman}}, \bibinfo {author} {\bibfnamefont {J.~H.}\
  \bibnamefont {Ng}}, \bibinfo {author} {\bibfnamefont {A.}~\bibnamefont
  {Nguyen}}, \bibinfo {author} {\bibfnamefont {M.}~\bibnamefont {Nguyen}},
  \bibinfo {author} {\bibfnamefont {M.~Y.}\ \bibnamefont {Niu}}, \bibinfo
  {author} {\bibfnamefont {S.}~\bibnamefont {Omonije}}, \bibinfo {author}
  {\bibfnamefont {A.}~\bibnamefont {Opremcak}}, \bibinfo {author}
  {\bibfnamefont {A.}~\bibnamefont {Petukhov}}, \bibinfo {author}
  {\bibfnamefont {R.}~\bibnamefont {Potter}}, \bibinfo {author} {\bibfnamefont
  {L.~P.}\ \bibnamefont {Pryadko}}, \bibinfo {author} {\bibfnamefont
  {C.}~\bibnamefont {Quintana}}, \bibinfo {author} {\bibfnamefont
  {C.}~\bibnamefont {Rocque}}, \bibinfo {author} {\bibfnamefont
  {P.}~\bibnamefont {Roushan}}, \bibinfo {author} {\bibfnamefont
  {N.}~\bibnamefont {Saei}}, \bibinfo {author} {\bibfnamefont {D.}~\bibnamefont
  {Sank}}, \bibinfo {author} {\bibfnamefont {K.}~\bibnamefont
  {Sankaragomathi}}, \bibinfo {author} {\bibfnamefont {K.~J.}\ \bibnamefont
  {Satzinger}}, \bibinfo {author} {\bibfnamefont {H.~F.}\ \bibnamefont
  {Schurkus}}, \bibinfo {author} {\bibfnamefont {C.}~\bibnamefont {Schuster}},
  \bibinfo {author} {\bibfnamefont {M.~J.}\ \bibnamefont {Shearn}}, \bibinfo
  {author} {\bibfnamefont {A.}~\bibnamefont {Shorter}}, \bibinfo {author}
  {\bibfnamefont {N.}~\bibnamefont {Shutty}}, \bibinfo {author} {\bibfnamefont
  {V.}~\bibnamefont {Shvarts}}, \bibinfo {author} {\bibfnamefont
  {J.}~\bibnamefont {Skruzny}}, \bibinfo {author} {\bibfnamefont {W.~C.}\
  \bibnamefont {Smith}}, \bibinfo {author} {\bibfnamefont {R.~D.}\ \bibnamefont
  {Somma}}, \bibinfo {author} {\bibfnamefont {G.}~\bibnamefont {Sterling}},
  \bibinfo {author} {\bibfnamefont {D.}~\bibnamefont {Strain}}, \bibinfo
  {author} {\bibfnamefont {M.}~\bibnamefont {Szalay}}, \bibinfo {author}
  {\bibfnamefont {D.}~\bibnamefont {Thor}}, \bibinfo {author} {\bibfnamefont
  {A.}~\bibnamefont {Torres}}, \bibinfo {author} {\bibfnamefont
  {G.}~\bibnamefont {Vidal}}, \bibinfo {author} {\bibfnamefont
  {B.}~\bibnamefont {Villalonga}}, \bibinfo {author} {\bibfnamefont
  {C.}~\bibnamefont {Vollgraff~Heidweiller}}, \bibinfo {author} {\bibfnamefont
  {T.}~\bibnamefont {White}}, \bibinfo {author} {\bibfnamefont {B.~W.~K.}\
  \bibnamefont {Woo}}, \bibinfo {author} {\bibfnamefont {C.}~\bibnamefont
  {Xing}}, \bibinfo {author} {\bibfnamefont {Z.~J.}\ \bibnamefont {Yao}},
  \bibinfo {author} {\bibfnamefont {P.}~\bibnamefont {Yeh}}, \bibinfo {author}
  {\bibfnamefont {J.}~\bibnamefont {Yoo}}, \bibinfo {author} {\bibfnamefont
  {G.}~\bibnamefont {Young}}, \bibinfo {author} {\bibfnamefont
  {A.}~\bibnamefont {Zalcman}}, \bibinfo {author} {\bibfnamefont
  {Y.}~\bibnamefont {Zhang}}, \bibinfo {author} {\bibfnamefont
  {N.}~\bibnamefont {Zhu}}, \bibinfo {author} {\bibfnamefont {N.}~\bibnamefont
  {Zobrist}}, \bibinfo {author} {\bibfnamefont {D.}~\bibnamefont {Bacon}},
  \bibinfo {author} {\bibfnamefont {S.}~\bibnamefont {Boixo}}, \bibinfo
  {author} {\bibfnamefont {Y.}~\bibnamefont {Chen}}, \bibinfo {author}
  {\bibfnamefont {J.}~\bibnamefont {Hilton}}, \bibinfo {author} {\bibfnamefont
  {J.}~\bibnamefont {Kelly}}, \bibinfo {author} {\bibfnamefont
  {E.}~\bibnamefont {Lucero}}, \bibinfo {author} {\bibfnamefont
  {A.}~\bibnamefont {Megrant}}, \bibinfo {author} {\bibfnamefont
  {H.}~\bibnamefont {Neven}}, \bibinfo {author} {\bibfnamefont
  {V.}~\bibnamefont {Smelyanskiy}}, \bibinfo {author} {\bibfnamefont
  {C.}~\bibnamefont {Gogolin}}, \bibinfo {author} {\bibfnamefont
  {R.}~\bibnamefont {Babbush}}, \ and\ \bibinfo {author} {\bibfnamefont
  {N.~C.}\ \bibnamefont {Rubin}},\ }\href {\doibase 10.1038/s41567-023-02240-y}
  {\bibfield  {journal} {\bibinfo  {journal} {Nat.~Phys.}\ }\textbf {\bibinfo
  {volume} {19}},\ \bibinfo {pages} {1787–1792} (\bibinfo {year}
  {2023})}\BibitemShut {NoStop}%
\bibitem [{\citenamefont {Liepuoniute}\ \emph {et~al.}(2024)\citenamefont
  {Liepuoniute}, \citenamefont {Motta}, \citenamefont {Pellegrini},
  \citenamefont {Rice}, \citenamefont {Gujarati}, \citenamefont {Gil},\ and\
  \citenamefont {Jones}}]{Liep24a}%
  \BibitemOpen
  \bibfield  {author} {\bibinfo {author} {\bibfnamefont {I.}~\bibnamefont
  {Liepuoniute}}, \bibinfo {author} {\bibfnamefont {M.}~\bibnamefont {Motta}},
  \bibinfo {author} {\bibfnamefont {T.}~\bibnamefont {Pellegrini}}, \bibinfo
  {author} {\bibfnamefont {J.~E.}\ \bibnamefont {Rice}}, \bibinfo {author}
  {\bibfnamefont {T.~P.}\ \bibnamefont {Gujarati}}, \bibinfo {author}
  {\bibfnamefont {S.}~\bibnamefont {Gil}}, \ and\ \bibinfo {author}
  {\bibfnamefont {G.~O.}\ \bibnamefont {Jones}},\ }\href {\doibase
  10.48550/ARXIV.2403.08107} {\enquote {\bibinfo {title} {Simulation of a
  diels-alder reaction on a quantum computer},}\ } (\bibinfo {year}
  {2024})\BibitemShut {NoStop}%
\bibitem [{\citenamefont {Robledo-Moreno}\ \emph {et~al.}(2024)\citenamefont
  {Robledo-Moreno}, \citenamefont {Motta}, \citenamefont {Haas}, \citenamefont
  {Javadi-Abhari}, \citenamefont {Jurcevic}, \citenamefont {Kirby},
  \citenamefont {Martiel}, \citenamefont {Sharma}, \citenamefont {Sharma},
  \citenamefont {Shirakawa}, \citenamefont {Sitdikov}, \citenamefont {Sun},
  \citenamefont {Sung}, \citenamefont {Takita}, \citenamefont {Tran},
  \citenamefont {Yunoki},\ and\ \citenamefont {Mezzacapo}}]{robl24a}%
  \BibitemOpen
  \bibfield  {author} {\bibinfo {author} {\bibfnamefont {J.}~\bibnamefont
  {Robledo-Moreno}}, \bibinfo {author} {\bibfnamefont {M.}~\bibnamefont
  {Motta}}, \bibinfo {author} {\bibfnamefont {H.}~\bibnamefont {Haas}},
  \bibinfo {author} {\bibfnamefont {A.}~\bibnamefont {Javadi-Abhari}}, \bibinfo
  {author} {\bibfnamefont {P.}~\bibnamefont {Jurcevic}}, \bibinfo {author}
  {\bibfnamefont {W.}~\bibnamefont {Kirby}}, \bibinfo {author} {\bibfnamefont
  {S.}~\bibnamefont {Martiel}}, \bibinfo {author} {\bibfnamefont
  {K.}~\bibnamefont {Sharma}}, \bibinfo {author} {\bibfnamefont
  {S.}~\bibnamefont {Sharma}}, \bibinfo {author} {\bibfnamefont
  {T.}~\bibnamefont {Shirakawa}}, \bibinfo {author} {\bibfnamefont
  {I.}~\bibnamefont {Sitdikov}}, \bibinfo {author} {\bibfnamefont {R.-Y.}\
  \bibnamefont {Sun}}, \bibinfo {author} {\bibfnamefont {K.~J.}\ \bibnamefont
  {Sung}}, \bibinfo {author} {\bibfnamefont {M.}~\bibnamefont {Takita}},
  \bibinfo {author} {\bibfnamefont {M.~C.}\ \bibnamefont {Tran}}, \bibinfo
  {author} {\bibfnamefont {S.}~\bibnamefont {Yunoki}}, \ and\ \bibinfo {author}
  {\bibfnamefont {A.}~\bibnamefont {Mezzacapo}},\ }\href {\doibase
  10.48550/ARXIV.2405.05068} {\enquote {\bibinfo {title} {Chemistry beyond
  exact solutions on a quantum-centric supercomputer},}\ } (\bibinfo {year}
  {2024})\BibitemShut {NoStop}%
\bibitem [{\citenamefont {Nagy}\ and\ \citenamefont {Jensen}(2017)}]{Nagy17}%
  \BibitemOpen
  \bibfield  {author} {\bibinfo {author} {\bibfnamefont {B.}~\bibnamefont
  {Nagy}}\ and\ \bibinfo {author} {\bibfnamefont {F.}~\bibnamefont {Jensen}},\
  }\href {\doibase 10.1002/9781119356059.ch3} {\bibfield  {journal} {\bibinfo
  {journal} {Reviews in Computational Chemistry}\ ,\ \bibinfo {pages}
  {93–149}} (\bibinfo {year} {2017})}\BibitemShut {NoStop}%
\bibitem [{not()}]{note-traditional}%
  \BibitemOpen
  \href@noop {} {}\bibinfo {note} {By ``traditional" the authors refer to
  quantum chemistry calculations performed on CPU and/or GPU-driven
  hardware.}\BibitemShut {Stop}%
\bibitem [{\citenamefont {Olsen}(2011)}]{olse11}%
  \BibitemOpen
  \bibfield  {author} {\bibinfo {author} {\bibfnamefont {J.}~\bibnamefont
  {Olsen}},\ }\href {\doibase 10.1002/qua.23107} {\bibfield  {journal}
  {\bibinfo  {journal} {Int. J. Quantum Chem.}\ }\textbf {\bibinfo {volume}
  {111}},\ \bibinfo {pages} {3267} (\bibinfo {year} {2011})}\BibitemShut
  {NoStop}%
\bibitem [{\citenamefont {Roos}\ \emph {et~al.}(2016)\citenamefont {Roos},
  \citenamefont {Lindh}, \citenamefont {Malmqvist}, \citenamefont {Veryazov},\
  and\ \citenamefont {Widmark}}]{roos16a}%
  \BibitemOpen
  \bibfield  {author} {\bibinfo {author} {\bibfnamefont {B.~O.}\ \bibnamefont
  {Roos}}, \bibinfo {author} {\bibfnamefont {R.}~\bibnamefont {Lindh}},
  \bibinfo {author} {\bibfnamefont {P.-{\AA}.}\ \bibnamefont {Malmqvist}},
  \bibinfo {author} {\bibfnamefont {V.}~\bibnamefont {Veryazov}}, \ and\
  \bibinfo {author} {\bibfnamefont {P.-O.}\ \bibnamefont {Widmark}},\ }\enquote
  {\bibinfo {title} {Multiconfigurational quantum chemistry},}\ in\ \href@noop
  {} {\emph {\bibinfo {booktitle} {Multiconfigurational Quantum Chemistry}}}\
  (\bibinfo  {publisher} {John Wiley \& Sons, Inc.},\ \bibinfo {year} {2016})\
  Chap.\ \bibinfo {chapter} {{CASPT2/CASSCF Applications}}, pp.\ \bibinfo
  {pages} {157--219}\BibitemShut {NoStop}%
\bibitem [{\citenamefont {Shee}\ \emph {et~al.}(2021)\citenamefont {Shee},
  \citenamefont {Loipersberger}, \citenamefont {Hait}, \citenamefont {Lee},\
  and\ \citenamefont {Head-Gordon}}]{shee21}%
  \BibitemOpen
  \bibfield  {author} {\bibinfo {author} {\bibfnamefont {J.}~\bibnamefont
  {Shee}}, \bibinfo {author} {\bibfnamefont {M.}~\bibnamefont {Loipersberger}},
  \bibinfo {author} {\bibfnamefont {D.}~\bibnamefont {Hait}}, \bibinfo {author}
  {\bibfnamefont {J.}~\bibnamefont {Lee}}, \ and\ \bibinfo {author}
  {\bibfnamefont {M.}~\bibnamefont {Head-Gordon}},\ }\href {\doibase
  10.1063/5.0047386} {\bibfield  {journal} {\bibinfo  {journal} {J. CHem.
  Phys.}\ }\textbf {\bibinfo {volume} {154}} (\bibinfo {year} {2021}),\
  10.1063/5.0047386}\BibitemShut {NoStop}%
\bibitem [{\citenamefont {Andersson}\ \emph {et~al.}(1990)\citenamefont
  {Andersson}, \citenamefont {Malmqvist}, \citenamefont {Roos}, \citenamefont
  {Sadlej},\ and\ \citenamefont {Wolinski}}]{Andersson1990}%
  \BibitemOpen
  \bibfield  {author} {\bibinfo {author} {\bibfnamefont {K.}~\bibnamefont
  {Andersson}}, \bibinfo {author} {\bibfnamefont {P.~A.}\ \bibnamefont
  {Malmqvist}}, \bibinfo {author} {\bibfnamefont {B.~O.}\ \bibnamefont {Roos}},
  \bibinfo {author} {\bibfnamefont {A.~J.}\ \bibnamefont {Sadlej}}, \ and\
  \bibinfo {author} {\bibfnamefont {K.}~\bibnamefont {Wolinski}},\ }\href
  {\doibase 10.1021/j100377a012} {\bibfield  {journal} {\bibinfo  {journal}
  {The Journal of Physical Chemistry}\ }\textbf {\bibinfo {volume} {94}},\
  \bibinfo {pages} {5483} (\bibinfo {year} {1990})}\BibitemShut {NoStop}%
\bibitem [{\citenamefont {Andersson}\ \emph
  {et~al.}(1992{\natexlab{a}})\citenamefont {Andersson}, \citenamefont
  {Malmqvist},\ and\ \citenamefont {Roos}}]{Roos1992}%
  \BibitemOpen
  \bibfield  {author} {\bibinfo {author} {\bibfnamefont {K.}~\bibnamefont
  {Andersson}}, \bibinfo {author} {\bibfnamefont {P.}~\bibnamefont
  {Malmqvist}}, \ and\ \bibinfo {author} {\bibfnamefont {B.~O.}\ \bibnamefont
  {Roos}},\ }\href {\doibase 10.1063/1.462209} {\bibfield  {journal} {\bibinfo
  {journal} {The Journal of Chemical Physics}\ }\textbf {\bibinfo {volume}
  {96}},\ \bibinfo {pages} {1218} (\bibinfo {year} {1992}{\natexlab{a}})},\
  \Eprint
  {http://arxiv.org/abs/https://pubs.aip.org/aip/jcp/article-pdf/96/2/1218/18996245/1218\_1\_online.pdf}
  {https://pubs.aip.org/aip/jcp/article-pdf/96/2/1218/18996245/1218\_1\_online.pdf}
  \BibitemShut {NoStop}%
\bibitem [{\citenamefont {Angeli}\ \emph
  {et~al.}(2001{\natexlab{a}})\citenamefont {Angeli}, \citenamefont
  {Cimiraglia},\ and\ \citenamefont {Malrieu}}]{ANGELI2001297}%
  \BibitemOpen
  \bibfield  {author} {\bibinfo {author} {\bibfnamefont {C.}~\bibnamefont
  {Angeli}}, \bibinfo {author} {\bibfnamefont {R.}~\bibnamefont {Cimiraglia}},
  \ and\ \bibinfo {author} {\bibfnamefont {J.-P.}\ \bibnamefont {Malrieu}},\
  }\href {\doibase https://doi.org/10.1016/S0009-2614(01)01303-3} {\bibfield
  {journal} {\bibinfo  {journal} {Chemical Physics Letters}\ }\textbf {\bibinfo
  {volume} {350}},\ \bibinfo {pages} {297} (\bibinfo {year}
  {2001}{\natexlab{a}})}\BibitemShut {NoStop}%
\bibitem [{\citenamefont {Angeli}\ \emph
  {et~al.}(2001{\natexlab{b}})\citenamefont {Angeli}, \citenamefont
  {Cimiraglia}, \citenamefont {Evangelisti}, \citenamefont {Leininger},\ and\
  \citenamefont {Malrieu}}]{malrieu2001}%
  \BibitemOpen
  \bibfield  {author} {\bibinfo {author} {\bibfnamefont {C.}~\bibnamefont
  {Angeli}}, \bibinfo {author} {\bibfnamefont {R.}~\bibnamefont {Cimiraglia}},
  \bibinfo {author} {\bibfnamefont {S.}~\bibnamefont {Evangelisti}}, \bibinfo
  {author} {\bibfnamefont {T.}~\bibnamefont {Leininger}}, \ and\ \bibinfo
  {author} {\bibfnamefont {J.-P.}\ \bibnamefont {Malrieu}},\ }\href {\doibase
  10.1063/1.1361246} {\bibfield  {journal} {\bibinfo  {journal} {The Journal of
  Chemical Physics}\ }\textbf {\bibinfo {volume} {114}},\ \bibinfo {pages}
  {10252} (\bibinfo {year} {2001}{\natexlab{b}})},\ \Eprint
  {http://arxiv.org/abs/https://pubs.aip.org/aip/jcp/article-pdf/114/23/10252/19105543/10252\_1\_online.pdf}
  {https://pubs.aip.org/aip/jcp/article-pdf/114/23/10252/19105543/10252\_1\_online.pdf}
  \BibitemShut {NoStop}%
\bibitem [{\citenamefont {Angeli}\ \emph
  {et~al.}(2002{\natexlab{a}})\citenamefont {Angeli}, \citenamefont
  {Cimiraglia},\ and\ \citenamefont {Malrieu}}]{jean-paul2002}%
  \BibitemOpen
  \bibfield  {author} {\bibinfo {author} {\bibfnamefont {C.}~\bibnamefont
  {Angeli}}, \bibinfo {author} {\bibfnamefont {R.}~\bibnamefont {Cimiraglia}},
  \ and\ \bibinfo {author} {\bibfnamefont {J.-P.}\ \bibnamefont {Malrieu}},\
  }\href {\doibase 10.1063/1.1515317} {\bibfield  {journal} {\bibinfo
  {journal} {The Journal of Chemical Physics}\ }\textbf {\bibinfo {volume}
  {117}},\ \bibinfo {pages} {9138} (\bibinfo {year} {2002}{\natexlab{a}})},\
  \Eprint
  {http://arxiv.org/abs/https://pubs.aip.org/aip/jcp/article-pdf/117/20/9138/19244630/9138\_1\_online.pdf}
  {https://pubs.aip.org/aip/jcp/article-pdf/117/20/9138/19244630/9138\_1\_online.pdf}
  \BibitemShut {NoStop}%
\bibitem [{\citenamefont {Sarkar}\ \emph {et~al.}(2022)\citenamefont {Sarkar},
  \citenamefont {Loos}, \citenamefont {Boggio-Pasqua},\ and\ \citenamefont
  {Jacquemin}}]{Sarkar2022}%
  \BibitemOpen
  \bibfield  {author} {\bibinfo {author} {\bibfnamefont {R.}~\bibnamefont
  {Sarkar}}, \bibinfo {author} {\bibfnamefont {P.-F.}\ \bibnamefont {Loos}},
  \bibinfo {author} {\bibfnamefont {M.}~\bibnamefont {Boggio-Pasqua}}, \ and\
  \bibinfo {author} {\bibfnamefont {D.}~\bibnamefont {Jacquemin}},\ }\href
  {\doibase 10.1021/acs.jctc.1c01197} {\bibfield  {journal} {\bibinfo
  {journal} {Journal of Chemical Theory and Computation}\ }\textbf {\bibinfo
  {volume} {18}},\ \bibinfo {pages} {2418} (\bibinfo {year} {2022})},\ \bibinfo
  {note} {pMID: 35333060},\ \Eprint
  {http://arxiv.org/abs/https://doi.org/10.1021/acs.jctc.1c01197}
  {https://doi.org/10.1021/acs.jctc.1c01197} \BibitemShut {NoStop}%
\bibitem [{\citenamefont {Stein}\ and\ \citenamefont {Reiher}(2016)}]{stei16a}%
  \BibitemOpen
  \bibfield  {author} {\bibinfo {author} {\bibfnamefont {C.~J.}\ \bibnamefont
  {Stein}}\ and\ \bibinfo {author} {\bibfnamefont {M.}~\bibnamefont {Reiher}},\
  }\href@noop {} {\bibfield  {journal} {\bibinfo  {journal} {J. Chem. Theory
  Comput.}\ }\textbf {\bibinfo {volume} {12}},\ \bibinfo {pages} {1760}
  (\bibinfo {year} {2016})}\BibitemShut {NoStop}%
\bibitem [{\citenamefont {Stein}\ and\ \citenamefont {Reiher}(2019)}]{stei19a}%
  \BibitemOpen
  \bibfield  {author} {\bibinfo {author} {\bibfnamefont {C.~J.}\ \bibnamefont
  {Stein}}\ and\ \bibinfo {author} {\bibfnamefont {M.}~\bibnamefont {Reiher}},\
  }\href@noop {} {\bibfield  {journal} {\bibinfo  {journal} {J. Comput. Chem.}\
  }\textbf {\bibinfo {volume} {40}},\ \bibinfo {pages} {2216} (\bibinfo {year}
  {2019})}\BibitemShut {NoStop}%
\bibitem [{\citenamefont {Sayfutyarova}\ \emph {et~al.}(2017)\citenamefont
  {Sayfutyarova}, \citenamefont {Sun}, \citenamefont {Chan},\ and\
  \citenamefont {Knizia}}]{sayf17a}%
  \BibitemOpen
  \bibfield  {author} {\bibinfo {author} {\bibfnamefont {E.~R.}\ \bibnamefont
  {Sayfutyarova}}, \bibinfo {author} {\bibfnamefont {Q.}~\bibnamefont {Sun}},
  \bibinfo {author} {\bibfnamefont {G.~K.-L.}\ \bibnamefont {Chan}}, \ and\
  \bibinfo {author} {\bibfnamefont {G.}~\bibnamefont {Knizia}},\ }\href
  {\doibase 10.1021/acs.jctc.7b00128} {\bibfield  {journal} {\bibinfo
  {journal} {J.~Chem.~Theory~Comput.}\ }\textbf {\bibinfo {volume} {13}},\
  \bibinfo {pages} {4063} (\bibinfo {year} {2017})}\BibitemShut {NoStop}%
\bibitem [{\citenamefont {Grimsley}\ \emph {et~al.}(2019)\citenamefont
  {Grimsley}, \citenamefont {Economou}, \citenamefont {Barnes},\ and\
  \citenamefont {Mayhall}}]{grim19a}%
  \BibitemOpen
  \bibfield  {author} {\bibinfo {author} {\bibfnamefont {H.~R.}\ \bibnamefont
  {Grimsley}}, \bibinfo {author} {\bibfnamefont {S.~E.}\ \bibnamefont
  {Economou}}, \bibinfo {author} {\bibfnamefont {E.}~\bibnamefont {Barnes}}, \
  and\ \bibinfo {author} {\bibfnamefont {N.~J.}\ \bibnamefont {Mayhall}},\
  }\href {\doibase 10.1038/s41467-019-10988-2} {\bibfield  {journal} {\bibinfo
  {journal} {Nat.~Commun.}\ }\textbf {\bibinfo {volume} {10}} (\bibinfo {year}
  {2019}),\ 10.1038/s41467-019-10988-2}\BibitemShut {NoStop}%
\bibitem [{\citenamefont {Fitzpatrick}\ \emph {et~al.}(2024)\citenamefont
  {Fitzpatrick}, \citenamefont {Nykänen}, \citenamefont {Talarico},
  \citenamefont {Lunghi}, \citenamefont {Maniscalco}, \citenamefont
  {García-Pérez},\ and\ \citenamefont
  {Knecht}}]{fitzpatrick2022selfconsistent}%
  \BibitemOpen
  \bibfield  {author} {\bibinfo {author} {\bibfnamefont {A.}~\bibnamefont
  {Fitzpatrick}}, \bibinfo {author} {\bibfnamefont {A.}~\bibnamefont
  {Nykänen}}, \bibinfo {author} {\bibfnamefont {N.~W.}\ \bibnamefont
  {Talarico}}, \bibinfo {author} {\bibfnamefont {A.}~\bibnamefont {Lunghi}},
  \bibinfo {author} {\bibfnamefont {S.}~\bibnamefont {Maniscalco}}, \bibinfo
  {author} {\bibfnamefont {G.}~\bibnamefont {García-Pérez}}, \ and\ \bibinfo
  {author} {\bibfnamefont {S.}~\bibnamefont {Knecht}},\ }\href {\doibase
  10.1021/acs.jpca.3c05882} {\bibfield  {journal} {\bibinfo  {journal} {J.
  Phys. Chem. A}\ }\textbf {\bibinfo {volume} {128}},\ \bibinfo {pages} {2843}
  (\bibinfo {year} {2024})}\BibitemShut {NoStop}%
\bibitem [{\citenamefont {Mahajan}\ \emph {et~al.}(2019)\citenamefont
  {Mahajan}, \citenamefont {Blunt}, \citenamefont {Sabzevari},\ and\
  \citenamefont {Sharma}}]{Mahajan2019-va}%
  \BibitemOpen
  \bibfield  {author} {\bibinfo {author} {\bibfnamefont {A.}~\bibnamefont
  {Mahajan}}, \bibinfo {author} {\bibfnamefont {N.~S.}\ \bibnamefont {Blunt}},
  \bibinfo {author} {\bibfnamefont {I.}~\bibnamefont {Sabzevari}}, \ and\
  \bibinfo {author} {\bibfnamefont {S.}~\bibnamefont {Sharma}},\ }\href
  {\doibase 10.1063/1.5128115} {\bibfield  {journal} {\bibinfo  {journal} {J.
  Chem. Phys.}\ }\textbf {\bibinfo {volume} {151}},\ \bibinfo {pages} {211102}
  (\bibinfo {year} {2019})}\BibitemShut {NoStop}%
\bibitem [{\citenamefont {Blunt}\ \emph {et~al.}(2020)\citenamefont {Blunt},
  \citenamefont {Mahajan},\ and\ \citenamefont {Sharma}}]{Blunt2020-uj}%
  \BibitemOpen
  \bibfield  {author} {\bibinfo {author} {\bibfnamefont {N.~S.}\ \bibnamefont
  {Blunt}}, \bibinfo {author} {\bibfnamefont {A.}~\bibnamefont {Mahajan}}, \
  and\ \bibinfo {author} {\bibfnamefont {S.}~\bibnamefont {Sharma}},\ }\href
  {\doibase 10.1063/5.0023353} {\bibfield  {journal} {\bibinfo  {journal} {J.
  Chem. Phys.}\ }\textbf {\bibinfo {volume} {153}},\ \bibinfo {pages} {164120}
  (\bibinfo {year} {2020})}\BibitemShut {NoStop}%
\bibitem [{\citenamefont {Kollmar}\ \emph {et~al.}(2021)\citenamefont
  {Kollmar}, \citenamefont {Sivalingam}, \citenamefont {Guo},\ and\
  \citenamefont {Neese}}]{Kollmar2021}%
  \BibitemOpen
  \bibfield  {author} {\bibinfo {author} {\bibfnamefont {C.}~\bibnamefont
  {Kollmar}}, \bibinfo {author} {\bibfnamefont {K.}~\bibnamefont {Sivalingam}},
  \bibinfo {author} {\bibfnamefont {Y.}~\bibnamefont {Guo}}, \ and\ \bibinfo
  {author} {\bibfnamefont {F.}~\bibnamefont {Neese}},\ }\href {\doibase
  10.1063/5.0072129} {\bibfield  {journal} {\bibinfo  {journal} {J. Chem.
  Phys.}\ }\textbf {\bibinfo {volume} {155}} (\bibinfo {year} {2021}),\
  10.1063/5.0072129}\BibitemShut {NoStop}%
\bibitem [{\citenamefont {Krompiec}\ and\ \citenamefont
  {Ramo}(2022)}]{Krompiec2022}%
  \BibitemOpen
  \bibfield  {author} {\bibinfo {author} {\bibfnamefont {M.}~\bibnamefont
  {Krompiec}}\ and\ \bibinfo {author} {\bibfnamefont {D.~M.}\ \bibnamefont
  {Ramo}},\ }\href {\doibase 10.48550/ARXIV.2210.05702} {\enquote {\bibinfo
  {title} {Strongly contracted n-electron valence state perturbation theory
  using reduced density matrices from a quantum computer},}\ } (\bibinfo {year}
  {2022})\BibitemShut {NoStop}%
\bibitem [{\citenamefont {Tammaro}\ \emph {et~al.}(2023)\citenamefont
  {Tammaro}, \citenamefont {Galli}, \citenamefont {Rice},\ and\ \citenamefont
  {Motta}}]{Tammaro2023}%
  \BibitemOpen
  \bibfield  {author} {\bibinfo {author} {\bibfnamefont {A.}~\bibnamefont
  {Tammaro}}, \bibinfo {author} {\bibfnamefont {D.~E.}\ \bibnamefont {Galli}},
  \bibinfo {author} {\bibfnamefont {J.~E.}\ \bibnamefont {Rice}}, \ and\
  \bibinfo {author} {\bibfnamefont {M.}~\bibnamefont {Motta}},\ }\href
  {\doibase 10.1021/acs.jpca.2c07653} {\bibfield  {journal} {\bibinfo
  {journal} {J. Phys. Chem. A}\ }\textbf {\bibinfo {volume} {127}},\ \bibinfo
  {pages} {817–827} (\bibinfo {year} {2023})}\BibitemShut {NoStop}%
\bibitem [{\citenamefont {Garc{\'\i}a-P{\'e}rez}\ \emph
  {et~al.}(2021)\citenamefont {Garc{\'\i}a-P{\'e}rez}, \citenamefont {Rossi},
  \citenamefont {Sokolov}, \citenamefont {Tacchino}, \citenamefont
  {Barkoutsos}, \citenamefont {Mazzola}, \citenamefont {Tavernelli},\ and\
  \citenamefont {Maniscalco}}]{garcia2021learning}%
  \BibitemOpen
  \bibfield  {author} {\bibinfo {author} {\bibfnamefont {G.}~\bibnamefont
  {Garc{\'\i}a-P{\'e}rez}}, \bibinfo {author} {\bibfnamefont {M.~A.}\
  \bibnamefont {Rossi}}, \bibinfo {author} {\bibfnamefont {B.}~\bibnamefont
  {Sokolov}}, \bibinfo {author} {\bibfnamefont {F.}~\bibnamefont {Tacchino}},
  \bibinfo {author} {\bibfnamefont {P.~K.}\ \bibnamefont {Barkoutsos}},
  \bibinfo {author} {\bibfnamefont {G.}~\bibnamefont {Mazzola}}, \bibinfo
  {author} {\bibfnamefont {I.}~\bibnamefont {Tavernelli}}, \ and\ \bibinfo
  {author} {\bibfnamefont {S.}~\bibnamefont {Maniscalco}},\ }\href {\doibase
  10.1103/PRXQuantum.2.040342} {\bibfield  {journal} {\bibinfo  {journal} {PRX
  Quantum}\ }\textbf {\bibinfo {volume} {2}},\ \bibinfo {pages} {040342}
  (\bibinfo {year} {2021})}\BibitemShut {NoStop}%
\bibitem [{\citenamefont {Glos}\ \emph {et~al.}(2022)\citenamefont {Glos},
  \citenamefont {Nyk{\"a}nen}, \citenamefont {Borrelli}, \citenamefont
  {Maniscalco}, \citenamefont {Rossi}, \citenamefont {Zimbor{\'a}s},\ and\
  \citenamefont {Garc{\'\i}a-P{\'e}rez}}]{glos22a}%
  \BibitemOpen
  \bibfield  {author} {\bibinfo {author} {\bibfnamefont {A.}~\bibnamefont
  {Glos}}, \bibinfo {author} {\bibfnamefont {A.}~\bibnamefont {Nyk{\"a}nen}},
  \bibinfo {author} {\bibfnamefont {E.-M.}\ \bibnamefont {Borrelli}}, \bibinfo
  {author} {\bibfnamefont {S.}~\bibnamefont {Maniscalco}}, \bibinfo {author}
  {\bibfnamefont {M.~A.}\ \bibnamefont {Rossi}}, \bibinfo {author}
  {\bibfnamefont {Z.}~\bibnamefont {Zimbor{\'a}s}}, \ and\ \bibinfo {author}
  {\bibfnamefont {G.}~\bibnamefont {Garc{\'\i}a-P{\'e}rez}},\ }\href {\doibase
  10.48550/ARXIV.2208.07817} {\enquote {\bibinfo {title} {Adaptive povm
  implementations and measurement error mitigation strategies for near-term
  quantum devices},}\ } (\bibinfo {year} {2022})\BibitemShut {NoStop}%
\bibitem [{\citenamefont {Nyk\"{a}nen}\ \emph {et~al.}(2022)\citenamefont
  {Nyk\"{a}nen}, \citenamefont {Rossi}, \citenamefont {Borrelli}, \citenamefont
  {Maniscalco},\ and\ \citenamefont {Garc{\'i}a-P{\'e}rez}}]{nyka22a}%
  \BibitemOpen
  \bibfield  {author} {\bibinfo {author} {\bibfnamefont {A.}~\bibnamefont
  {Nyk\"{a}nen}}, \bibinfo {author} {\bibfnamefont {M.~A.~C.}\ \bibnamefont
  {Rossi}}, \bibinfo {author} {\bibfnamefont {E.-M.}\ \bibnamefont {Borrelli}},
  \bibinfo {author} {\bibfnamefont {S.}~\bibnamefont {Maniscalco}}, \ and\
  \bibinfo {author} {\bibfnamefont {G.}~\bibnamefont {Garc{\'i}a-P{\'e}rez}},\
  }\href {\doibase 10.48550/ARXIV.2212.09719} {\enquote {\bibinfo {title}
  {Mitigating the measurement overhead of {ADAPT-VQE} with optimised
  informationally complete generalised measurements},}\ } (\bibinfo {year}
  {2022}),\ \bibinfo {note} {arXiv: 2212.09719}\BibitemShut {NoStop}%
\bibitem [{\citenamefont {Stein}\ \emph {et~al.}(2016)\citenamefont {Stein},
  \citenamefont {von Burg},\ and\ \citenamefont {Reiher}}]{stei16b}%
  \BibitemOpen
  \bibfield  {author} {\bibinfo {author} {\bibfnamefont {C.~J.}\ \bibnamefont
  {Stein}}, \bibinfo {author} {\bibfnamefont {V.}~\bibnamefont {von Burg}}, \
  and\ \bibinfo {author} {\bibfnamefont {M.}~\bibnamefont {Reiher}},\
  }\href@noop {} {\bibfield  {journal} {\bibinfo  {journal} {J. Chem. Theory
  Comput.}\ }\textbf {\bibinfo {volume} {12}},\ \bibinfo {pages} {3764}
  (\bibinfo {year} {2016})}\BibitemShut {NoStop}%
\bibitem [{\citenamefont {Roos}\ \emph {et~al.}(1982)\citenamefont {Roos},
  \citenamefont {Linse}, \citenamefont {Siegbahn},\ and\ \citenamefont
  {Blomberg}}]{Roos1982-th}%
  \BibitemOpen
  \bibfield  {author} {\bibinfo {author} {\bibfnamefont {B.~O.}\ \bibnamefont
  {Roos}}, \bibinfo {author} {\bibfnamefont {P.}~\bibnamefont {Linse}},
  \bibinfo {author} {\bibfnamefont {P.~E.~M.}\ \bibnamefont {Siegbahn}}, \ and\
  \bibinfo {author} {\bibfnamefont {M.~R.~A.}\ \bibnamefont {Blomberg}},\
  }\href {\doibase 10.1016/0301-0104(82)88019-1} {\bibfield  {journal}
  {\bibinfo  {journal} {Chem. Phys.}\ }\textbf {\bibinfo {volume} {66}},\
  \bibinfo {pages} {197} (\bibinfo {year} {1982})}\BibitemShut {NoStop}%
\bibitem [{\citenamefont {Andersson}\ \emph
  {et~al.}(1992{\natexlab{b}})\citenamefont {Andersson}, \citenamefont
  {Malmqvist},\ and\ \citenamefont {Roos}}]{Andersson1992-rc}%
  \BibitemOpen
  \bibfield  {author} {\bibinfo {author} {\bibfnamefont {K.}~\bibnamefont
  {Andersson}}, \bibinfo {author} {\bibfnamefont {P.-{\AA}.}\ \bibnamefont
  {Malmqvist}}, \ and\ \bibinfo {author} {\bibfnamefont {B.~O.}\ \bibnamefont
  {Roos}},\ }\href {\doibase 10.1063/1.462209} {\bibfield  {journal} {\bibinfo
  {journal} {J. Chem. Phys.}\ }\textbf {\bibinfo {volume} {96}},\ \bibinfo
  {pages} {1218} (\bibinfo {year} {1992}{\natexlab{b}})}\BibitemShut {NoStop}%
\bibitem [{\citenamefont {Angeli}\ \emph
  {et~al.}(2001{\natexlab{c}})\citenamefont {Angeli}, \citenamefont
  {Cimiraglia}, \citenamefont {Evangelisti}, \citenamefont {Leininger},\ and\
  \citenamefont {Malrieu}}]{Angeli2001}%
  \BibitemOpen
  \bibfield  {author} {\bibinfo {author} {\bibfnamefont {C.}~\bibnamefont
  {Angeli}}, \bibinfo {author} {\bibfnamefont {R.}~\bibnamefont {Cimiraglia}},
  \bibinfo {author} {\bibfnamefont {S.}~\bibnamefont {Evangelisti}}, \bibinfo
  {author} {\bibfnamefont {T.}~\bibnamefont {Leininger}}, \ and\ \bibinfo
  {author} {\bibfnamefont {J.-P.}\ \bibnamefont {Malrieu}},\ }\href {\doibase
  10.1063/1.1361246} {\bibfield  {journal} {\bibinfo  {journal} {The Journal of
  Chemical Physics}\ }\textbf {\bibinfo {volume} {114}},\ \bibinfo {pages}
  {10252} (\bibinfo {year} {2001}{\natexlab{c}})}\BibitemShut {NoStop}%
\bibitem [{\citenamefont {Angeli}\ \emph
  {et~al.}(2001{\natexlab{d}})\citenamefont {Angeli}, \citenamefont
  {Cimiraglia},\ and\ \citenamefont {Malrieu}}]{Angeli2001_2}%
  \BibitemOpen
  \bibfield  {author} {\bibinfo {author} {\bibfnamefont {C.}~\bibnamefont
  {Angeli}}, \bibinfo {author} {\bibfnamefont {R.}~\bibnamefont {Cimiraglia}},
  \ and\ \bibinfo {author} {\bibfnamefont {J.-P.}\ \bibnamefont {Malrieu}},\
  }\href {\doibase 10.1016/s0009-2614(01)01303-3} {\bibfield  {journal}
  {\bibinfo  {journal} {Chemical Physics Letters}\ }\textbf {\bibinfo {volume}
  {350}},\ \bibinfo {pages} {297} (\bibinfo {year}
  {2001}{\natexlab{d}})}\BibitemShut {NoStop}%
\bibitem [{\citenamefont {Angeli}\ \emph
  {et~al.}(2002{\natexlab{b}})\citenamefont {Angeli}, \citenamefont
  {Cimiraglia},\ and\ \citenamefont {Malrieu}}]{Angeli2002-xw}%
  \BibitemOpen
  \bibfield  {author} {\bibinfo {author} {\bibfnamefont {C.}~\bibnamefont
  {Angeli}}, \bibinfo {author} {\bibfnamefont {R.}~\bibnamefont {Cimiraglia}},
  \ and\ \bibinfo {author} {\bibfnamefont {J.-P.}\ \bibnamefont {Malrieu}},\
  }\href {\doibase 10.1063/1.1515317} {\bibfield  {journal} {\bibinfo
  {journal} {J. Chem. Phys.}\ }\textbf {\bibinfo {volume} {117}},\ \bibinfo
  {pages} {9138} (\bibinfo {year} {2002}{\natexlab{b}})}\BibitemShut {NoStop}%
\bibitem [{\citenamefont {Sokolov}\ and\ \citenamefont
  {Chan}(2016)}]{Sokolov2016}%
  \BibitemOpen
  \bibfield  {author} {\bibinfo {author} {\bibfnamefont {A.~Y.}\ \bibnamefont
  {Sokolov}}\ and\ \bibinfo {author} {\bibfnamefont {G.~K.-L.}\ \bibnamefont
  {Chan}},\ }\href {\doibase 10.1063/1.4941606} {\bibfield  {journal} {\bibinfo
   {journal} {The Journal of Chemical Physics}\ }\textbf {\bibinfo {volume}
  {144}},\ \bibinfo {pages} {064102} (\bibinfo {year} {2016})}\BibitemShut
  {NoStop}%
\bibitem [{\citenamefont {Sokolov}\ \emph {et~al.}(2017)\citenamefont
  {Sokolov}, \citenamefont {Guo}, \citenamefont {Ronca},\ and\ \citenamefont
  {Chan}}]{Sokolov2017}%
  \BibitemOpen
  \bibfield  {author} {\bibinfo {author} {\bibfnamefont {A.~Y.}\ \bibnamefont
  {Sokolov}}, \bibinfo {author} {\bibfnamefont {S.}~\bibnamefont {Guo}},
  \bibinfo {author} {\bibfnamefont {E.}~\bibnamefont {Ronca}}, \ and\ \bibinfo
  {author} {\bibfnamefont {G.~K.-L.}\ \bibnamefont {Chan}},\ }\href {\doibase
  10.1063/1.4986975} {\bibfield  {journal} {\bibinfo  {journal} {The Journal of
  Chemical Physics}\ }\textbf {\bibinfo {volume} {146}},\ \bibinfo {pages}
  {244102} (\bibinfo {year} {2017})}\BibitemShut {NoStop}%
\bibitem [{\citenamefont {Pulay}(2011)}]{Pulay2011}%
  \BibitemOpen
  \bibfield  {author} {\bibinfo {author} {\bibfnamefont {P.}~\bibnamefont
  {Pulay}},\ }\href {\doibase 10.1002/qua.23052} {\bibfield  {journal}
  {\bibinfo  {journal} {International Journal of Quantum Chemistry}\ }\textbf
  {\bibinfo {volume} {111}},\ \bibinfo {pages} {3273} (\bibinfo {year}
  {2011})}\BibitemShut {NoStop}%
\bibitem [{\citenamefont {Dyall}(1995)}]{Dyall1995-ay}%
  \BibitemOpen
  \bibfield  {author} {\bibinfo {author} {\bibfnamefont {K.~G.}\ \bibnamefont
  {Dyall}},\ }\href {\doibase 10.1063/1.469539} {\bibfield  {journal} {\bibinfo
   {journal} {J. Chem. Phys.}\ }\textbf {\bibinfo {volume} {102}},\ \bibinfo
  {pages} {4909} (\bibinfo {year} {1995})}\BibitemShut {NoStop}%
\bibitem [{\citenamefont {Kitaev}(1995)}]{kita95a}%
  \BibitemOpen
  \bibfield  {author} {\bibinfo {author} {\bibfnamefont {A.~Y.}\ \bibnamefont
  {Kitaev}},\ }\href {\doibase 10.48550/ARXIV.QUANT-PH/9511026} {\enquote
  {\bibinfo {title} {Quantum measurements and the abelian stabilizer
  problem},}\ } (\bibinfo {year} {1995})\BibitemShut {NoStop}%
\bibitem [{\citenamefont {Guther}\ \emph {et~al.}(2020)\citenamefont {Guther},
  \citenamefont {Anderson}, \citenamefont {Blunt}, \citenamefont {Bogdanov},
  \citenamefont {Cleland}, \citenamefont {Dattani}, \citenamefont {Dobrautz},
  \citenamefont {Ghanem}, \citenamefont {Jeszenszki}, \citenamefont
  {Liebermann}, \citenamefont {Manni}, \citenamefont {Lozovoi}, \citenamefont
  {Luo}, \citenamefont {Ma}, \citenamefont {Merz}, \citenamefont {Overy},
  \citenamefont {Rampp}, \citenamefont {Samanta}, \citenamefont {Schwarz},
  \citenamefont {Shepherd}, \citenamefont {Smart}, \citenamefont {Vitale},
  \citenamefont {Weser}, \citenamefont {Booth},\ and\ \citenamefont
  {Alavi}}]{guth20a}%
  \BibitemOpen
  \bibfield  {author} {\bibinfo {author} {\bibfnamefont {K.}~\bibnamefont
  {Guther}}, \bibinfo {author} {\bibfnamefont {R.~J.}\ \bibnamefont
  {Anderson}}, \bibinfo {author} {\bibfnamefont {N.~S.}\ \bibnamefont {Blunt}},
  \bibinfo {author} {\bibfnamefont {N.~A.}\ \bibnamefont {Bogdanov}}, \bibinfo
  {author} {\bibfnamefont {D.}~\bibnamefont {Cleland}}, \bibinfo {author}
  {\bibfnamefont {N.}~\bibnamefont {Dattani}}, \bibinfo {author} {\bibfnamefont
  {W.}~\bibnamefont {Dobrautz}}, \bibinfo {author} {\bibfnamefont
  {K.}~\bibnamefont {Ghanem}}, \bibinfo {author} {\bibfnamefont
  {P.}~\bibnamefont {Jeszenszki}}, \bibinfo {author} {\bibfnamefont
  {N.}~\bibnamefont {Liebermann}}, \bibinfo {author} {\bibfnamefont {G.~L.}\
  \bibnamefont {Manni}}, \bibinfo {author} {\bibfnamefont {A.~Y.}\ \bibnamefont
  {Lozovoi}}, \bibinfo {author} {\bibfnamefont {H.}~\bibnamefont {Luo}},
  \bibinfo {author} {\bibfnamefont {D.}~\bibnamefont {Ma}}, \bibinfo {author}
  {\bibfnamefont {F.}~\bibnamefont {Merz}}, \bibinfo {author} {\bibfnamefont
  {C.}~\bibnamefont {Overy}}, \bibinfo {author} {\bibfnamefont
  {M.}~\bibnamefont {Rampp}}, \bibinfo {author} {\bibfnamefont {P.~K.}\
  \bibnamefont {Samanta}}, \bibinfo {author} {\bibfnamefont {L.~R.}\
  \bibnamefont {Schwarz}}, \bibinfo {author} {\bibfnamefont {J.~J.}\
  \bibnamefont {Shepherd}}, \bibinfo {author} {\bibfnamefont {S.~D.}\
  \bibnamefont {Smart}}, \bibinfo {author} {\bibfnamefont {E.}~\bibnamefont
  {Vitale}}, \bibinfo {author} {\bibfnamefont {O.}~\bibnamefont {Weser}},
  \bibinfo {author} {\bibfnamefont {G.~H.}\ \bibnamefont {Booth}}, \ and\
  \bibinfo {author} {\bibfnamefont {A.}~\bibnamefont {Alavi}},\ }\href
  {\doibase 10.1063/5.0005754} {\bibfield  {journal} {\bibinfo  {journal}
  {J.~Chem.~Phys.}\ }\textbf {\bibinfo {volume} {153}},\ \bibinfo {pages}
  {034107} (\bibinfo {year} {2020})}\BibitemShut {NoStop}%
\bibitem [{\citenamefont {Tilly}\ \emph {et~al.}(2022)\citenamefont {Tilly},
  \citenamefont {Chen}, \citenamefont {Cao}, \citenamefont {Picozzi},
  \citenamefont {Setia}, \citenamefont {Li}, \citenamefont {Grant},
  \citenamefont {Wossnig}, \citenamefont {Rungger}, \citenamefont {Booth},\
  and\ \citenamefont {Tennyson}}]{till22a}%
  \BibitemOpen
  \bibfield  {author} {\bibinfo {author} {\bibfnamefont {J.}~\bibnamefont
  {Tilly}}, \bibinfo {author} {\bibfnamefont {H.}~\bibnamefont {Chen}},
  \bibinfo {author} {\bibfnamefont {S.}~\bibnamefont {Cao}}, \bibinfo {author}
  {\bibfnamefont {D.}~\bibnamefont {Picozzi}}, \bibinfo {author} {\bibfnamefont
  {K.}~\bibnamefont {Setia}}, \bibinfo {author} {\bibfnamefont
  {Y.}~\bibnamefont {Li}}, \bibinfo {author} {\bibfnamefont {E.}~\bibnamefont
  {Grant}}, \bibinfo {author} {\bibfnamefont {L.}~\bibnamefont {Wossnig}},
  \bibinfo {author} {\bibfnamefont {I.}~\bibnamefont {Rungger}}, \bibinfo
  {author} {\bibfnamefont {G.~H.}\ \bibnamefont {Booth}}, \ and\ \bibinfo
  {author} {\bibfnamefont {J.}~\bibnamefont {Tennyson}},\ }\href {\doibase
  10.1016/j.physrep.2022.08.003} {\bibfield  {journal} {\bibinfo  {journal}
  {Phys.~Rep.}\ }\textbf {\bibinfo {volume} {986}},\ \bibinfo {pages} {1}
  (\bibinfo {year} {2022})}\BibitemShut {NoStop}%
\bibitem [{\citenamefont {Fedorov}\ \emph {et~al.}(2022)\citenamefont
  {Fedorov}, \citenamefont {Peng}, \citenamefont {Govind},\ and\ \citenamefont
  {Alexeev}}]{fedo22a}%
  \BibitemOpen
  \bibfield  {author} {\bibinfo {author} {\bibfnamefont {D.~A.}\ \bibnamefont
  {Fedorov}}, \bibinfo {author} {\bibfnamefont {B.}~\bibnamefont {Peng}},
  \bibinfo {author} {\bibfnamefont {N.}~\bibnamefont {Govind}}, \ and\ \bibinfo
  {author} {\bibfnamefont {Y.}~\bibnamefont {Alexeev}},\ }\href {\doibase
  10.1186/s41313-021-00032-6} {\bibfield  {journal} {\bibinfo  {journal}
  {Mat.~Theory}\ }\textbf {\bibinfo {volume} {6}},\ \bibinfo {pages} {1}
  (\bibinfo {year} {2022})}\BibitemShut {NoStop}%
\bibitem [{\citenamefont {G\"{u}nther}\ \emph {et~al.}(2023)\citenamefont
  {G\"{u}nther}, \citenamefont {Baiardi}, \citenamefont {Reiher},\ and\
  \citenamefont {Christandl}}]{Gunther2023}%
  \BibitemOpen
  \bibfield  {author} {\bibinfo {author} {\bibfnamefont {J.}~\bibnamefont
  {G\"{u}nther}}, \bibinfo {author} {\bibfnamefont {A.}~\bibnamefont
  {Baiardi}}, \bibinfo {author} {\bibfnamefont {M.}~\bibnamefont {Reiher}}, \
  and\ \bibinfo {author} {\bibfnamefont {M.}~\bibnamefont {Christandl}},\
  }\href {\doibase 10.48550/ARXIV.2308.16873} {\enquote {\bibinfo {title} {More
  quantum chemistry with fewer qubits},}\ } (\bibinfo {year} {2023}),\ \bibinfo
  {note} {arXiv: 2308.16873}\BibitemShut {NoStop}%
\bibitem [{\citenamefont {Kutzelnigg}\ \emph {et~al.}(2010)\citenamefont
  {Kutzelnigg}, \citenamefont {Shamasundar},\ and\ \citenamefont
  {Mukherjee}}]{kutz10a}%
  \BibitemOpen
  \bibfield  {author} {\bibinfo {author} {\bibfnamefont {W.}~\bibnamefont
  {Kutzelnigg}}, \bibinfo {author} {\bibfnamefont {K.}~\bibnamefont
  {Shamasundar}}, \ and\ \bibinfo {author} {\bibfnamefont {D.}~\bibnamefont
  {Mukherjee}},\ }\href@noop {} {\bibfield  {journal} {\bibinfo  {journal}
  {Mol. Phys.}\ }\textbf {\bibinfo {volume} {108}},\ \bibinfo {pages} {433}
  (\bibinfo {year} {2010})}\BibitemShut {NoStop}%
\bibitem [{\citenamefont {McClean}\ \emph {et~al.}(2017)\citenamefont
  {McClean}, \citenamefont {Kimchi-Schwartz}, \citenamefont {Carter},\ and\
  \citenamefont {de~Jong}}]{mcclean2017}%
  \BibitemOpen
  \bibfield  {author} {\bibinfo {author} {\bibfnamefont {J.~R.}\ \bibnamefont
  {McClean}}, \bibinfo {author} {\bibfnamefont {M.~E.}\ \bibnamefont
  {Kimchi-Schwartz}}, \bibinfo {author} {\bibfnamefont {J.}~\bibnamefont
  {Carter}}, \ and\ \bibinfo {author} {\bibfnamefont {W.~A.}\ \bibnamefont
  {de~Jong}},\ }\href {\doibase 10.1103/PhysRevA.95.042308} {\bibfield
  {journal} {\bibinfo  {journal} {Phys. Rev. A}\ }\textbf {\bibinfo {volume}
  {95}},\ \bibinfo {pages} {042308} (\bibinfo {year} {2017})}\BibitemShut
  {NoStop}%
\bibitem [{\citenamefont {Gonthier}\ \emph {et~al.}(2022)\citenamefont
  {Gonthier}, \citenamefont {Radin}, \citenamefont {Buda}, \citenamefont
  {Doskocil}, \citenamefont {Abuan},\ and\ \citenamefont
  {Romero}}]{Gonthier2022}%
  \BibitemOpen
  \bibfield  {author} {\bibinfo {author} {\bibfnamefont {J.~F.}\ \bibnamefont
  {Gonthier}}, \bibinfo {author} {\bibfnamefont {M.~D.}\ \bibnamefont {Radin}},
  \bibinfo {author} {\bibfnamefont {C.}~\bibnamefont {Buda}}, \bibinfo {author}
  {\bibfnamefont {E.~J.}\ \bibnamefont {Doskocil}}, \bibinfo {author}
  {\bibfnamefont {C.~M.}\ \bibnamefont {Abuan}}, \ and\ \bibinfo {author}
  {\bibfnamefont {J.}~\bibnamefont {Romero}},\ }\href {\doibase
  10.1103/PhysRevResearch.4.033154} {\bibfield  {journal} {\bibinfo  {journal}
  {Phys. Rev. Res.}\ }\textbf {\bibinfo {volume} {4}},\ \bibinfo {pages}
  {033154} (\bibinfo {year} {2022})}\BibitemShut {NoStop}%
\bibitem [{\citenamefont {Izmaylov}\ \emph {et~al.}(2020)\citenamefont
  {Izmaylov}, \citenamefont {Yen}, \citenamefont {Lang},\ and\ \citenamefont
  {Verteletskyi}}]{Izmaylov2019}%
  \BibitemOpen
  \bibfield  {author} {\bibinfo {author} {\bibfnamefont {A.~F.}\ \bibnamefont
  {Izmaylov}}, \bibinfo {author} {\bibfnamefont {T.-C.}\ \bibnamefont {Yen}},
  \bibinfo {author} {\bibfnamefont {R.~A.}\ \bibnamefont {Lang}}, \ and\
  \bibinfo {author} {\bibfnamefont {V.}~\bibnamefont {Verteletskyi}},\ }\href
  {\doibase 10.1021/acs.jctc.9b00791} {\bibfield  {journal} {\bibinfo
  {journal} {Journal of Chemical Theory and Computation}\ }\textbf {\bibinfo
  {volume} {16}},\ \bibinfo {pages} {190} (\bibinfo {year} {2020})},\ \bibinfo
  {note} {pMID: 31747266},\ \Eprint
  {http://arxiv.org/abs/https://doi.org/10.1021/acs.jctc.9b00791}
  {https://doi.org/10.1021/acs.jctc.9b00791} \BibitemShut {NoStop}%
\bibitem [{\citenamefont {Zhao}\ \emph {et~al.}(2020)\citenamefont {Zhao},
  \citenamefont {Tranter}, \citenamefont {Kirby}, \citenamefont {Ung},
  \citenamefont {Miyake},\ and\ \citenamefont {Love}}]{Zhao2020}%
  \BibitemOpen
  \bibfield  {author} {\bibinfo {author} {\bibfnamefont {A.}~\bibnamefont
  {Zhao}}, \bibinfo {author} {\bibfnamefont {A.}~\bibnamefont {Tranter}},
  \bibinfo {author} {\bibfnamefont {W.~M.}\ \bibnamefont {Kirby}}, \bibinfo
  {author} {\bibfnamefont {S.~F.}\ \bibnamefont {Ung}}, \bibinfo {author}
  {\bibfnamefont {A.}~\bibnamefont {Miyake}}, \ and\ \bibinfo {author}
  {\bibfnamefont {P.~J.}\ \bibnamefont {Love}},\ }\href {\doibase
  10.1103/PhysRevA.101.062322} {\bibfield  {journal} {\bibinfo  {journal}
  {Phys. Rev. A}\ }\textbf {\bibinfo {volume} {101}},\ \bibinfo {pages}
  {062322} (\bibinfo {year} {2020})}\BibitemShut {NoStop}%
\bibitem [{\citenamefont {Huggins}\ \emph {et~al.}(2021)\citenamefont
  {Huggins}, \citenamefont {McArdle}, \citenamefont {O'Brien}, \citenamefont
  {Lee}, \citenamefont {Rubin}, \citenamefont {Boixo}, \citenamefont {Whaley},
  \citenamefont {Babbush},\ and\ \citenamefont {McClean}}]{Huggins2021}%
  \BibitemOpen
  \bibfield  {author} {\bibinfo {author} {\bibfnamefont {W.~J.}\ \bibnamefont
  {Huggins}}, \bibinfo {author} {\bibfnamefont {S.}~\bibnamefont {McArdle}},
  \bibinfo {author} {\bibfnamefont {T.~E.}\ \bibnamefont {O'Brien}}, \bibinfo
  {author} {\bibfnamefont {J.}~\bibnamefont {Lee}}, \bibinfo {author}
  {\bibfnamefont {N.~C.}\ \bibnamefont {Rubin}}, \bibinfo {author}
  {\bibfnamefont {S.}~\bibnamefont {Boixo}}, \bibinfo {author} {\bibfnamefont
  {K.~B.}\ \bibnamefont {Whaley}}, \bibinfo {author} {\bibfnamefont
  {R.}~\bibnamefont {Babbush}}, \ and\ \bibinfo {author} {\bibfnamefont
  {J.~R.}\ \bibnamefont {McClean}},\ }\href {\doibase
  10.1103/PhysRevX.11.041036} {\bibfield  {journal} {\bibinfo  {journal} {Phys.
  Rev. X}\ }\textbf {\bibinfo {volume} {11}},\ \bibinfo {pages} {041036}
  (\bibinfo {year} {2021})}\BibitemShut {NoStop}%
\bibitem [{\citenamefont {Crawford}\ \emph {et~al.}(2021)\citenamefont
  {Crawford}, \citenamefont {van Straaten}, \citenamefont {Wang}, \citenamefont
  {Parks}, \citenamefont {Campbell},\ and\ \citenamefont
  {Brierley}}]{Crawford2021}%
  \BibitemOpen
  \bibfield  {author} {\bibinfo {author} {\bibfnamefont {O.}~\bibnamefont
  {Crawford}}, \bibinfo {author} {\bibfnamefont {B.}~\bibnamefont {van
  Straaten}}, \bibinfo {author} {\bibfnamefont {D.}~\bibnamefont {Wang}},
  \bibinfo {author} {\bibfnamefont {T.}~\bibnamefont {Parks}}, \bibinfo
  {author} {\bibfnamefont {E.}~\bibnamefont {Campbell}}, \ and\ \bibinfo
  {author} {\bibfnamefont {S.}~\bibnamefont {Brierley}},\ }\href {\doibase
  10.22331/q-2021-01-20-385} {\bibfield  {journal} {\bibinfo  {journal}
  {Quantum}\ }\textbf {\bibinfo {volume} {5}},\ \bibinfo {pages} {385}
  (\bibinfo {year} {2021})}\BibitemShut {NoStop}%
\bibitem [{\citenamefont {Verteletskyi}\ \emph {et~al.}(2020)\citenamefont
  {Verteletskyi}, \citenamefont {Yen},\ and\ \citenamefont
  {Izmaylov}}]{Verteletskyi2020}%
  \BibitemOpen
  \bibfield  {author} {\bibinfo {author} {\bibfnamefont {V.}~\bibnamefont
  {Verteletskyi}}, \bibinfo {author} {\bibfnamefont {T.-C.}\ \bibnamefont
  {Yen}}, \ and\ \bibinfo {author} {\bibfnamefont {A.~F.}\ \bibnamefont
  {Izmaylov}},\ }\href {\doibase 10.1063/1.5141458} {\bibfield  {journal}
  {\bibinfo  {journal} {The Journal of Chemical Physics}\ }\textbf {\bibinfo
  {volume} {152}},\ \bibinfo {pages} {124114} (\bibinfo {year} {2020})},\
  \Eprint
  {http://arxiv.org/abs/https://pubs.aip.org/aip/jcp/article-pdf/doi/10.1063/1.5141458/15573883/124114\_1\_online.pdf}
  {https://pubs.aip.org/aip/jcp/article-pdf/doi/10.1063/1.5141458/15573883/124114\_1\_online.pdf}
  \BibitemShut {NoStop}%
\bibitem [{\citenamefont {Fischer}\ \emph {et~al.}(2022)\citenamefont
  {Fischer}, \citenamefont {Miller}, \citenamefont {Tacchino}, \citenamefont
  {Barkoutsos}, \citenamefont {Egger},\ and\ \citenamefont
  {Tavernelli}}]{Fischer2022}%
  \BibitemOpen
  \bibfield  {author} {\bibinfo {author} {\bibfnamefont {L.~E.}\ \bibnamefont
  {Fischer}}, \bibinfo {author} {\bibfnamefont {D.}~\bibnamefont {Miller}},
  \bibinfo {author} {\bibfnamefont {F.}~\bibnamefont {Tacchino}}, \bibinfo
  {author} {\bibfnamefont {P.~K.}\ \bibnamefont {Barkoutsos}}, \bibinfo
  {author} {\bibfnamefont {D.~J.}\ \bibnamefont {Egger}}, \ and\ \bibinfo
  {author} {\bibfnamefont {I.}~\bibnamefont {Tavernelli}},\ }\href {\doibase
  10.1103/PhysRevResearch.4.033027} {\bibfield  {journal} {\bibinfo  {journal}
  {Phys. Rev. Res.}\ }\textbf {\bibinfo {volume} {4}},\ \bibinfo {pages}
  {033027} (\bibinfo {year} {2022})}\BibitemShut {NoStop}%
\bibitem [{\citenamefont {Sun}\ \emph {et~al.}(2017)\citenamefont {Sun},
  \citenamefont {Berkelbach}, \citenamefont {Blunt}, \citenamefont {Booth},
  \citenamefont {Guo}, \citenamefont {Li}, \citenamefont {Liu}, \citenamefont
  {McClain}, \citenamefont {Sayfutyarova}, \citenamefont {Sharma},
  \citenamefont {Wouters},\ and\ \citenamefont {Chan}}]{sunq17b}%
  \BibitemOpen
  \bibfield  {author} {\bibinfo {author} {\bibfnamefont {Q.}~\bibnamefont
  {Sun}}, \bibinfo {author} {\bibfnamefont {T.~C.}\ \bibnamefont {Berkelbach}},
  \bibinfo {author} {\bibfnamefont {N.~S.}\ \bibnamefont {Blunt}}, \bibinfo
  {author} {\bibfnamefont {G.~H.}\ \bibnamefont {Booth}}, \bibinfo {author}
  {\bibfnamefont {S.}~\bibnamefont {Guo}}, \bibinfo {author} {\bibfnamefont
  {Z.}~\bibnamefont {Li}}, \bibinfo {author} {\bibfnamefont {J.}~\bibnamefont
  {Liu}}, \bibinfo {author} {\bibfnamefont {J.~D.}\ \bibnamefont {McClain}},
  \bibinfo {author} {\bibfnamefont {E.~R.}\ \bibnamefont {Sayfutyarova}},
  \bibinfo {author} {\bibfnamefont {S.}~\bibnamefont {Sharma}}, \bibinfo
  {author} {\bibfnamefont {S.}~\bibnamefont {Wouters}}, \ and\ \bibinfo
  {author} {\bibfnamefont {G.~K.-L.}\ \bibnamefont {Chan}},\ }\href {\doibase
  10.1002/wcms.1340} {\bibfield  {journal} {\bibinfo  {journal}
  {{WIREs~Comput.~Mol.~Sci}}\ }\textbf {\bibinfo {volume} {8}} (\bibinfo {year}
  {2017}),\ 10.1002/wcms.1340}\BibitemShut {NoStop}%
\bibitem [{\citenamefont {Sun}\ \emph {et~al.}(2020)\citenamefont {Sun},
  \citenamefont {Zhang}, \citenamefont {Banerjee}, \citenamefont {Bao},
  \citenamefont {Barbry}, \citenamefont {Blunt}, \citenamefont {Bogdanov},
  \citenamefont {Booth}, \citenamefont {Chen}, \citenamefont {Cui},
  \citenamefont {Eriksen}, \citenamefont {Gao}, \citenamefont {Guo},
  \citenamefont {Hermann}, \citenamefont {Hermes}, \citenamefont {Koh},
  \citenamefont {Koval}, \citenamefont {Lehtola}, \citenamefont {Li},
  \citenamefont {Liu}, \citenamefont {Mardirossian}, \citenamefont {McClain},
  \citenamefont {Motta}, \citenamefont {Mussard}, \citenamefont {Pham},
  \citenamefont {Pulkin}, \citenamefont {Purwanto}, \citenamefont {Robinson},
  \citenamefont {Ronca}, \citenamefont {Sayfutyarova}, \citenamefont
  {Scheurer}, \citenamefont {Schurkus}, \citenamefont {Smith}, \citenamefont
  {Sun}, \citenamefont {Sun}, \citenamefont {Upadhyay}, \citenamefont {Wagner},
  \citenamefont {Wang}, \citenamefont {White}, \citenamefont {Whitfield},
  \citenamefont {Williamson}, \citenamefont {Wouters}, \citenamefont {Yang},
  \citenamefont {Yu}, \citenamefont {Zhu}, \citenamefont {Berkelbach},
  \citenamefont {Sharma}, \citenamefont {Sokolov},\ and\ \citenamefont
  {Chan}}]{sunq20a}%
  \BibitemOpen
  \bibfield  {author} {\bibinfo {author} {\bibfnamefont {Q.}~\bibnamefont
  {Sun}}, \bibinfo {author} {\bibfnamefont {X.}~\bibnamefont {Zhang}}, \bibinfo
  {author} {\bibfnamefont {S.}~\bibnamefont {Banerjee}}, \bibinfo {author}
  {\bibfnamefont {P.}~\bibnamefont {Bao}}, \bibinfo {author} {\bibfnamefont
  {M.}~\bibnamefont {Barbry}}, \bibinfo {author} {\bibfnamefont {N.~S.}\
  \bibnamefont {Blunt}}, \bibinfo {author} {\bibfnamefont {N.~A.}\ \bibnamefont
  {Bogdanov}}, \bibinfo {author} {\bibfnamefont {G.~H.}\ \bibnamefont {Booth}},
  \bibinfo {author} {\bibfnamefont {J.}~\bibnamefont {Chen}}, \bibinfo {author}
  {\bibfnamefont {Z.-H.}\ \bibnamefont {Cui}}, \bibinfo {author} {\bibfnamefont
  {J.~J.}\ \bibnamefont {Eriksen}}, \bibinfo {author} {\bibfnamefont
  {Y.}~\bibnamefont {Gao}}, \bibinfo {author} {\bibfnamefont {S.}~\bibnamefont
  {Guo}}, \bibinfo {author} {\bibfnamefont {J.}~\bibnamefont {Hermann}},
  \bibinfo {author} {\bibfnamefont {M.~R.}\ \bibnamefont {Hermes}}, \bibinfo
  {author} {\bibfnamefont {K.}~\bibnamefont {Koh}}, \bibinfo {author}
  {\bibfnamefont {P.}~\bibnamefont {Koval}}, \bibinfo {author} {\bibfnamefont
  {S.}~\bibnamefont {Lehtola}}, \bibinfo {author} {\bibfnamefont
  {Z.}~\bibnamefont {Li}}, \bibinfo {author} {\bibfnamefont {J.}~\bibnamefont
  {Liu}}, \bibinfo {author} {\bibfnamefont {N.}~\bibnamefont {Mardirossian}},
  \bibinfo {author} {\bibfnamefont {J.~D.}\ \bibnamefont {McClain}}, \bibinfo
  {author} {\bibfnamefont {M.}~\bibnamefont {Motta}}, \bibinfo {author}
  {\bibfnamefont {B.}~\bibnamefont {Mussard}}, \bibinfo {author} {\bibfnamefont
  {H.~Q.}\ \bibnamefont {Pham}}, \bibinfo {author} {\bibfnamefont
  {A.}~\bibnamefont {Pulkin}}, \bibinfo {author} {\bibfnamefont
  {W.}~\bibnamefont {Purwanto}}, \bibinfo {author} {\bibfnamefont {P.~J.}\
  \bibnamefont {Robinson}}, \bibinfo {author} {\bibfnamefont {E.}~\bibnamefont
  {Ronca}}, \bibinfo {author} {\bibfnamefont {E.~R.}\ \bibnamefont
  {Sayfutyarova}}, \bibinfo {author} {\bibfnamefont {M.}~\bibnamefont
  {Scheurer}}, \bibinfo {author} {\bibfnamefont {H.~F.}\ \bibnamefont
  {Schurkus}}, \bibinfo {author} {\bibfnamefont {J.~E.~T.}\ \bibnamefont
  {Smith}}, \bibinfo {author} {\bibfnamefont {C.}~\bibnamefont {Sun}}, \bibinfo
  {author} {\bibfnamefont {S.-N.}\ \bibnamefont {Sun}}, \bibinfo {author}
  {\bibfnamefont {S.}~\bibnamefont {Upadhyay}}, \bibinfo {author}
  {\bibfnamefont {L.~K.}\ \bibnamefont {Wagner}}, \bibinfo {author}
  {\bibfnamefont {X.}~\bibnamefont {Wang}}, \bibinfo {author} {\bibfnamefont
  {A.}~\bibnamefont {White}}, \bibinfo {author} {\bibfnamefont {J.~D.}\
  \bibnamefont {Whitfield}}, \bibinfo {author} {\bibfnamefont {M.~J.}\
  \bibnamefont {Williamson}}, \bibinfo {author} {\bibfnamefont
  {S.}~\bibnamefont {Wouters}}, \bibinfo {author} {\bibfnamefont
  {J.}~\bibnamefont {Yang}}, \bibinfo {author} {\bibfnamefont {J.~M.}\
  \bibnamefont {Yu}}, \bibinfo {author} {\bibfnamefont {T.}~\bibnamefont
  {Zhu}}, \bibinfo {author} {\bibfnamefont {T.~C.}\ \bibnamefont {Berkelbach}},
  \bibinfo {author} {\bibfnamefont {S.}~\bibnamefont {Sharma}}, \bibinfo
  {author} {\bibfnamefont {A.~Y.}\ \bibnamefont {Sokolov}}, \ and\ \bibinfo
  {author} {\bibfnamefont {G.~K.-L.}\ \bibnamefont {Chan}},\ }\href {\doibase
  10.1063/5.0006074} {\bibfield  {journal} {\bibinfo  {journal}
  {J.~Chem.~Phys.}\ }\textbf {\bibinfo {volume} {153}},\ \bibinfo {pages}
  {024109} (\bibinfo {year} {2020})}\BibitemShut {NoStop}%
\bibitem [{aur(2024)}]{aurora}%
  \BibitemOpen
  \href@noop {} {} (\bibinfo {year} {2024}),\ \bibinfo {note} {\textsc{Aurora},
  Algorithmiq Ltd.}\BibitemShut {Stop}%
\bibitem [{\citenamefont {{Dunning Jr.}}(1989)}]{dunn89}%
  \BibitemOpen
  \bibfield  {author} {\bibinfo {author} {\bibfnamefont {T.~H.}\ \bibnamefont
  {{Dunning Jr.}}},\ }\href {\doibase 10.1063/1.456153} {\bibfield  {journal}
  {\bibinfo  {journal} {J. Chem. Phys.}\ }\textbf {\bibinfo {volume} {90}},\
  \bibinfo {pages} {1007} (\bibinfo {year} {1989})}\BibitemShut {NoStop}%
\bibitem [{\citenamefont {Nyk\"{a}nen}\ \emph {et~al.}(2024)\citenamefont
  {Nyk\"{a}nen}, \citenamefont {Thiessen}, \citenamefont {Borrelli},
  \citenamefont {Krishna}, \citenamefont {Knecht},\ and\ \citenamefont {Pavo{\u
  s}evi{\'c}}}]{pavo24b}%
  \BibitemOpen
  \bibfield  {author} {\bibinfo {author} {\bibfnamefont {A.}~\bibnamefont
  {Nyk\"{a}nen}}, \bibinfo {author} {\bibfnamefont {L.}~\bibnamefont
  {Thiessen}}, \bibinfo {author} {\bibfnamefont {E.-M.}\ \bibnamefont
  {Borrelli}}, \bibinfo {author} {\bibfnamefont {V.}~\bibnamefont {Krishna}},
  \bibinfo {author} {\bibfnamefont {S.}~\bibnamefont {Knecht}}, \ and\ \bibinfo
  {author} {\bibfnamefont {F.}~\bibnamefont {Pavo{\u s}evi{\'c}}},\ }\href
  {\doibase 10.48550/ARXIV.2404.16149} {\enquote {\bibinfo {title}
  {$\delta$adapt-vqe: Toward accurate calculation of excitation energies on
  quantum computers for bodipy molecules with application in photodynamic
  therapy},}\ } (\bibinfo {year} {2024}),\ \bibinfo {note} {arXiv:
  2404.16149}\BibitemShut {NoStop}%
\end{thebibliography}%
\end{document}